\documentclass[%
 reprint,
 twocolumn,
 superscriptaddress,
 showpacs,
 preprintnumbers,
 amsmath,
 amssymb,
 aps,
 prd,
 floatfix,
]{revtex4-1}

\usepackage{epsfig}
\usepackage{multirow}
\usepackage{hyperref} 

\newcommand{\bra}{\langle}
\newcommand{\ket}{\rangle}

\newcommand{\Tr}{\mbox{Tr}}

\newcommand{\half}{\frac{1}{2}}

\newcommand{\hmu}{\hat{\mu}}
\newcommand{\hnu}{\hat{\nu}}
\newcommand{\ipr}{i^\prime}

\newcommand{\hhtime}{\hat{4}}

\newcommand{\xv}{{\mathbf x}}

\newcommand{\Tpc}{T_{\rm pc}}

\newcommand{\be}{\begin{equation}}
\newcommand{\ee}{\end{equation}}
\newcommand{\bea}{\begin{eqnarray}}
\newcommand{\eea}{\end{eqnarray}}
\newcommand{\bean}{\begin{eqnarray*}}
\newcommand{\eean}{\end{eqnarray*}}
\newcommand{\nn}{\nonumber}

\begin{document}

\title{Properties of the QCD thermal transition with $N_f=2+1$ flavours of Wilson quark}

\author{G.~Aarts}
\email[Corresponding author: ]{g.aarts@swansea.ac.uk}
\affiliation{Department of Physics, College of Science, Swansea University, Swansea SA2 8PP, United Kingdom}

\author{C.~Allton}
\affiliation{Department of Physics, College of Science, Swansea University, Swansea SA2 8PP, United Kingdom}

\author{J.~Glesaaen}
\affiliation{Department of Physics, College of Science, Swansea University, Swansea SA2 8PP, United Kingdom}

\author{S.~Hands}
\affiliation{Department of Physics, College of Science, Swansea University, Swansea SA2 8PP, United Kingdom}

\author{B.~J{\"a}ger}
\affiliation{CP$^3$-Origins \& Danish IAS, Department of Mathematics and Computer Science, University of Southern Denmark, 5230 Odense M, Denmark}

\author{S.~Kim}
\affiliation{Department of Physics, Sejong University, Seoul 143-747, Korea}

\author{M.~P.~Lombardo}
\affiliation{INFN, Sezione di Firenze, 50019 Sesto Fiorentino (FI), Italy}

\author{A.~A.~Nikolaev}
\email[Corresponding author: ]{aleksandr.nikolaev@swansea.ac.uk}
\affiliation{Department of Physics, College of Science, Swansea University, Swansea SA2 8PP, United Kingdom}

\author{S.~M.~Ryan}
\affiliation{School of Mathematics and Hamilton Mathematics Institute, Trinity College, Dublin 2, Ireland}

\author{J.-I.~Skullerud}
\affiliation{School of Mathematics and Hamilton Mathematics Institute, Trinity College, Dublin 2, Ireland}
\affiliation{Department of Theoretical Physics, National University of Ireland Maynooth, Maynooth, County Kildare, Ireland}

\author{L.-K.~Wu}
\affiliation{Faculty of Science, Jiangsu University, Zhenjiang, 212013 \& Key Laboratory of Quark and Lepton Physics (MOE), Central China Normal University, Wuhan 430079, China}

\date{July 7, 2020}

\begin{abstract}
We study properties of the thermal transition in QCD, using anisotropic, fixed-scale lattice simulations with $N_f = 2+1$ flavours of Wilson fermion. Observables are compared for two values of the pion mass, focusing on chiral properties. Results are presented for the Polyakov loop, various susceptibilities, the chiral condensate and its susceptibility, and the onset of parity doubling in the light and strange baryonic sector.
\end{abstract}


\maketitle


\section{Introduction}
\label{sec:intro}

Mapping out the QCD phase diagram remains one of the outstanding challenges in the theory of the strong interactions. By now, it is well established that the transition along the temperature axis, at vanishing baryon density, is a crossover~\cite{Aoki:2006we}. This result has been obtained, and confirmed, using simulations of lattice QCD with physical quark masses in the continuum limit~\cite{Borsanyi:2010bp,Borsanyi:2010cj,Bazavov:2014pvz}. It is expected that the transition becomes a proper phase transition for quarks lighter than those in nature, reflecting the chiral symmetry of massless quarks. The manner in which this occurs depends on the way the chiral limit is taken, {\it e.g.} by considering $N_f=3$ degenerate quark flavours, or instead the $N_f=2+1$ case, with the strange quark mass fixed at its physical value. In the latter situation, a possible scenario is that the chiral transition is second order for exactly massless light quarks only, but a crossover for nonzero quark masses~\cite{Kogut:1982fn,Pisarski:1983ms}. This aspect of the QCD thermal transition, including the value of the transition temperature in the chiral limit, is currently an active area of study, see, {\it e.g.}, Refs.~\cite{Ding:2019prx,Braun:2020ada,Braguta:2019yci}. Most lattice studies, including those mentioned above, have been carried out using the staggered fermion formulation. It is important to investigate properties of the crossover with alternative fermion formulations, such as Wilson quarks, which avoid any potential uncertainty with this approach, {\it e.g.} with regard to the rooting of staggered fermions and taste symmetry violations. This provides one motivation for the work presented in this paper.

Besides the phase structure, many questions arise related to spectroscopy, {\it i.e.} the behaviour of hadrons as the temperature of the hadronic gas is increased to approach and then exceed the crossover temperature, turning the system into a quark-gluon plasma (QGP). This is highly relevant for heavy-ion phenomenology, where the in-medium modification and melting of hadrons provides an important characterisation of the QGP. In a sequence of papers some of us have studied this question, for heavy-quark bound states (bottomonium)~\cite{Aarts:2010ek,Aarts:2011sm,Aarts:2013kaa,Aarts:2014cda}, hidden and open charm~\cite{Kelly:2018hsi}, and positive- and negative-parity light baryons~\cite{Aarts:2015mma,Aarts:2017rrl} and hyperons~\cite{Aarts:2018glk}. In these studies we used Wilson fermions, for which there is a clear practical motivation: all time slices are available for spectroscopic analysis, avoiding the staggering present in temporal correlators obtained using the staggered formulation.  In addition, we use anisotropic lattices, with $a_\tau /a_s \ll 1$ (here $a_\tau$ and $a_s$ are the temporal and spatial lattice spacing respectively),  to further increase the number of data points  in the temporal direction available for analysis. The studies listed above have been obtained at a single lattice spacing, using light quarks that are heavier than in nature, while the strange quark takes its physical value. In order to improve on this, one has to systematically reduce the lattice spacing and the two light quark masses. Due to the anisotropy, this is a nontrivial endeavour as a tuning of the bare parameters at $T=0$ (gauge and fermion anisotropies, light and strange quark masses) is required for each value of the lattice spacing and quark masses,  done in such a way that the anisotropy is kept approximately constant. As a next step in this programme, we present here a new set of ensembles at eleven different temperatures for lighter quarks,  reducing the pion mass from approximately 384 MeV (employed in Refs.~\  \cite{Aarts:2014cda,Kelly:2018hsi,Aarts:2015mma,Aarts:2017rrl,Aarts:2018glk}) to 236 MeV, keeping the lattice spacing unchanged. In order to embark on a spectroscopic analysis of these ensembles, it is necessary to characterise them from a thermodynamic viewpoint and determine the properties of the thermal crossover.  This is the second motivation of this study.

In the remainder of this introduction, we discuss several other works that have employed $N_f=2+1$ Wilson quarks to investigate QCD at nonzero temperature, for comparison. As mentioned above, none of these studies have used physical quark masses and taken the continuum limit simultaneously, mostly due to the inherent cost in simulating Wilson fermions over staggered ones.
We note that all studies described below, including ours, use the fixed-scale approach, in which the temperature  $T=1/(a_\tau N_\tau)$ is varied by changing $N_\tau$ at fixed $a_\tau$. The benefit is that it is straightforward  to generate and compare ensembles at different temperatures, without the need to change the bare parameters, once the ensembles at $T=0$ have been tuned. This should be contrasted with the fixed-temperature approach, where the main goal is the extrapolation to the continuum limit, obtained by varying the lattice spacing and temporal extent of the lattice simultaneously, such that the temperature is kept fixed. 
In Refs.~\cite{Borsanyi:2012uq,Borsanyi:2015waa} the Budapest-Wuppertal group studied $N_f=2+1$ QCD thermodynamics on isotropic  lattices, while taking the continuum limit using two, three or four values of the lattice spacing. The pion masses were approximately 545, 440 and 285 MeV. As the pion mass is reduced, the pseudocritical temperature is seen to decrease, but no estimates for its value are given. The WHOT collaboration, in a series of papers~\cite{Umeda:2012er,Taniguchi:2016ofw,Taniguchi:2020mgg}, has studied $N_f=2+1$ QCD thermodynamics using gradient flow.  They employ isotropic lattices at a single lattice spacing, with a pion heavier than in nature. Preliminary results at the physical point are given in Ref.~\cite{Kanaya:2019okb}. The final related work we mention here employs twisted-mass fermions with $N_f=2+1+1$ flavours, including at the physical point, at a single lattice spacing~\cite{Burger:2018fvb,Kotov:2020hzm}. We come back to those results later on in the paper. We emphasise that all papers mentioned above use isotropic lattices.

This paper is organised as follows. In the following section, we introduce the new ensembles and make a brief comparison between these and the previous ones. The Polyakov loop and heavy-quark entropy are discussed in Sec.~\ref{sec:pol}. Susceptibilities related to quark number are analysed in Sec.~\ref{sec:sus}. Sec.~\ref{sec:chiral} gives results on the chiral condensate and its susceptibility. Results for parity doubling in light baryonic channels as a sign of chiral symmetry restoration are presented in Sec.~\ref{sec:parity}. A comparison of the various results for the pseudocritical temperature is finally given in Sec.~\ref{sec:summary}. Appendix~\ref{app:action} contains details of the lattice action, the parameter choices and the code used. Preliminary results have been presented in Refs.~\cite{Aarts:2018haw,Aarts:2019hrg}.


\section{Finite-temperature ensembles}
\label{sec:ensembles}

We employ the anisotropic lattice formulation introduced by the Hadron Spectrum Collaboration and use the same bare gauge and fermion anisotropies and bare sea quark masses as employed in their extensive spectroscopy programme, see for example Refs.~\cite{Wilson:2019wfr,Cheung:2016bym} and references therein. In brief, we employ a Symanzik-improved gauge action and a Wilson tadpole-improved clover fermion action, with stout-smeared links. Further details of the action are given in Appendix~\ref{app:action} of this paper; the full details of the action and the parameter tuning strategy were described in Refs.~\cite{Edwards:2008ja,Lin:2008pr}.
In our previous work \cite{Aarts:2014nba,Aarts:2014cda,Kelly:2018hsi,Aarts:2015mma,Aarts:2017rrl,Aarts:2018glk}, the $N_f=2+1$ Generation 2 (Gen2) ensembles corresponded to a physical strange quark mass and a bare light quark mass of $a_\tau m_l=-0.0840$, yielding a pion mass of $m_\pi=384(4)$ MeV (see Table \ref{tab:lattice_spacings}).
The latter was determined from exponential fits to a $3\times 3$ matrix of Gaussian-smeared correlation functions~\cite{Edwards:2008ja}. The pion mass quoted more recently by the Hadron Spectrum Collaboration is $m_\pi=391$ MeV, determined from $\pi\pi$ P-wave scattering, using distillation and a large basis of interpolating operators on multiple lattice volumes~\cite{Dudek:2012gj,Wilson:2019wfr}. We will use the value of $m_\pi=384(4)$ MeV to indicate the Gen2 results in the plots below. The target anisotropy is 3.5; the renormalised anisotropy $\xi$ is given in Table \ref{tab:lattice_spacings}.

The new $N_f=2+1$ Generation 2L (Gen2L, L for light) ensembles have the same
physical strange quark mass and lighter (degenerate) up and down quark masses, with a bare mass of $a_\tau m_l=-0.0860$. Following the Hadron Spectrum Collaboration as before, this corresponds to a pion mass of $m_\pi=236(2)$ MeV~\cite{Wilson:2019wfr}.
The Gen2L ensembles introduced here allow a study of the light quark mass dependence, with almost all other parameters unchanged in the simulation. The spatial lattice volume is increased ($N_s=24\rightarrow 32$) to ensure a large enough physical volume ($m_\pi L>4$), and the anisotropy, $\xi=a_s/a_\tau$, measured from the pion dispersion relation on the lowest temperature ensembles is approximately the same~\cite{Wilson:2019wfr}.
A comparison of the two ensembles is given in Table~\ref{tab:lattice_spacings}. 

\begin{table}[t]
  \begin{center}
    \begin{tabular}[t]{|c||c|c|}
	\hline
	& Gen2 & Gen2L \\
	\hline
	$a_\tau$ [fm] & 0.0350(2) & 0.0330(2)  \\
	\; $a_\tau^{-1}$ [GeV] \; & 5.63(4) & 5.997(34)  \\
	$\xi=a_s/a_\tau$ &  3.444(6)  & 3.453(6)  \\
	$a_s$ [fm] & \; 0.1205(8) \; & \; 0.1136(6) \;  \\
	$N_s$   &   24 & 32  \\
	$m_\pi$ [MeV] & 384(4) & 236(2) \\
	$m_\pi L$ & 5.63 & 4.36 \\
	\hline
  \end{tabular}
  \end{center}
\caption{Comparison of Generation 2 and 2L ensembles. The temporal lattice spacing is determined using the mass of $\Omega$ baryon. $\xi$ is the renormalised anisotropy, determined via the slope of the pion dispersion relation.}
  \label{tab:lattice_spacings}
\end{table}

\begin{table}[t]
  \begin{center}
     \begin{tabular}[t]{|c|c|c|c|c|}
	\hline
 	\multicolumn{5}{|c|}{Generation 2, $24^3 \times N_\tau$} \\
    	\hline
	$N_\tau$ & \; $T$ [MeV] \;  & \; $T/T_c$ \; & \; $N_{\rm cfg}$ \; & \; $N_{\rm stoch}$ \; \\
 	\hline
	\; 128$^*$ \; & 44 & 0.24 & 305 & 100 \\
 	48$^\dagger$ & 117 & 0.63 & 251 & 1200 \\
 	40 & 141 & 0.76 & 502  & 800 \\
 	36 & 156 & 0.84 & 501  & 400 \\
 	32 & 176 & 0.95 & 1000 & 400 \\
 	28 & 201 & 1.09 & 1001 & 400 \\ 
 	24 & 235 & 1.27 & 1002 & 100 \\
 	20 & 281 & 1.52 & 1000 & 100 \\
 	16 & 352 & 1.90 & 1000 & 100 \\ 
	\hline
    \end{tabular}
   
   \vspace*{0.5cm} 
   
    \begin{tabular}[t]{|c|c|c|c|}
	\hline
	\multicolumn{4}{|c|}{Generation 2L, $32^3 \times N_\tau$} \\
	\hline
    	$N_\tau$ & \; $T$ [MeV] \; & \; $N_{\rm cfg}$ \; & \; $N_{\rm stoch}$ \;  \\
	\hline
	\; 256$^*$ \;  & 23 & 750 & $-$  \\
 	128 & 47 & 1024 & 400     \\
   	64 & 94  & 1041 & 1600    \\
   	56 & 107 & 1042 & 1600    \\
   	48 & 125 & 1123 & 1200   \\
   	40 & 150 & 1102 & 1200    \\ 
   	36 & 167 & 1119 & 800     \\
   	32 & 187 & 1090 & 400     \\
   	28 & 214 & 1031 & 400    \\ 
  	24 & 250 & 1016 & 400    \\
   	20 & 300 & 1030 & 100    \\
   	16 & 375 & 1102 & 100    \\
   	12 & 500 & 1267 & $-$     \\
    	8  & 750 & 1048 & $-$     \\ 
	\hline
   \end{tabular}
    \end{center}
     \caption{Temporal extent, temperature in MeV, number of configurations, and number of Gaussian random vectors, used for susceptibilities for the ensembles of Generation 2 (above) and Generation 2L (below). The ensembles at the lowest temperatures, marked by an $^*$, were provided by HadSpec~\cite{Edwards:2008ja,Lin:2008pr} (Gen2), \cite{Wilson:2015dqa,Cheung:2016bym} (Gen2L). Ensembles marked with ``$-$'' were not used for results presented in this paper. The $N_\tau=48^\dagger$ Gen2 ensemble is on a spatial volume of $32^3$. }
    \label{tab:Gens}
\end{table}

In the fixed-scale approach, it is straightforward to generate ensembles at nonzero temperature, simply by changing the temporal extent $N_\tau$. Details of the finite-temperature ensembles are listed in Table~\ref{tab:Gens}. Here $N_{\rm cfg}$ refers to the number of independent configurations generated (after thermalisation) and  $N_{\rm stoch}$ to the number of Gaussian random vectors used for the computation of susceptibilities. The ensembles at the lowest temperatures, labelled with a $^*$, have been kindly provided by HadSpec, although we do not use the $N_\tau = 256$ ensemble in this work, and consider the $N_\tau = 128$ ensembles as the ``$T\approx 0$" ensemble. Since these ensembles satisfy $N_\tau>\xi N_s$, or $1/T>L$, it is indeed appropriate to consider them to be at zero temperature. For the sake of consistency of notation, we will assign a nominal temperature $T=1/(a_\tau N_\tau)$ to these ensembles in the following. The Gen2L ensembles at the two highest temperatures ($N_\tau = 12,\,8$) are not used either; since the temperatures are above 500 MeV, they do not provide additional information on the thermal transition. In Gen2, we include one ensemble on a $32^3$ volume, namely with $N_\tau=48$, to increase the number of available temperatures in the hadronic phase, in particular for the analysis of the chiral condensate and susceptibilities.

In the next sections we will present an overview of the crossover as inferred from the Polyakov loop and in particular from fermionic observables (susceptibilities, chiral condensate, baryon parity doubling). We note here that for Gen2 the pseudocritical temperature has already been  determined via the renormalised Polyakov loop and estimated to be $T^{P}_{\rm pc} = 185(4)$ MeV~\cite{Aarts:2014nba}. This value of $T_{\rm pc}$ is used in the third column of Table~\ref{tab:Gens} for Gen2, leading to four ensembles above and five below $\Tpc$. 
For Gen2L we will see that a reliable estimate for the pseudocritical temperature follows from the chiral condensate, with $T^{\bar\psi\psi}_{\rm pc}=162(1)$ MeV. Hence there are a sufficient number of ensembles in both the hadronic phase and the quark-gluon plasma to allow us to study the thermal transition in detail.


\section{Polyakov loop}
\label{sec:pol}

The Polyakov loop acts as an order parameter for the spontaneous breaking of centre symmetry at high temperature in Yang-Mills theory. In the presence of quarks, centre symmetry is explicitly broken and the Polyakov loop no longer plays this role. Nevertheless, it is often used as an indicator of the thermal transition, although its relevance is diminished as the simulated quarks becomes lighter~\cite{Borsanyi:2010bp,Borsanyi:2010cj,Bazavov:2014pvz}.

The Polyakov loop is defined, on a single configuration, as the trace of the product of the links in the temporal direction,
\be 
\label{eq:Polyakov_loop_per_conf}
P_\xv = \frac{1}{3} \Tr \prod_{\tau = 0}^{N_\tau-1} U_{(\tau, \xv), 4}.
\ee
Similarly, the conjugate Polyakov loop is given by $P_\xv^\dagger$. Their expectation values are real and directly related to the free energy of an infinitely heavy (anti-)quark, 
\be
L_{\rm bare} = \bra P_\xv \ket = e^{-F^{q}/T},\qquad L^c_{\rm bare} = \bra P_\xv^\dagger\ket = e^{-F^{\bar q}/T}.
\ee
The subscript `bare' is used to emphasise that these are unrenormalised. At vanishing baryon chemical potential, $F^{q}=F^{\bar q}$ and 
\be
L_{\rm bare}  L^c_{\rm bare} = \bra P_\xv \ket \bra P_\xv^\dagger \ket = e^{-2F^{q}/T}.
\ee
This expression is useful when analysing simulations, since the imaginary parts of $L_{\rm bare}$ and $L^c_{\rm bare}$ both fluctuate around zero.

\begin{figure}[t]
\begin{center}
\includegraphics[width=0.48\textwidth]{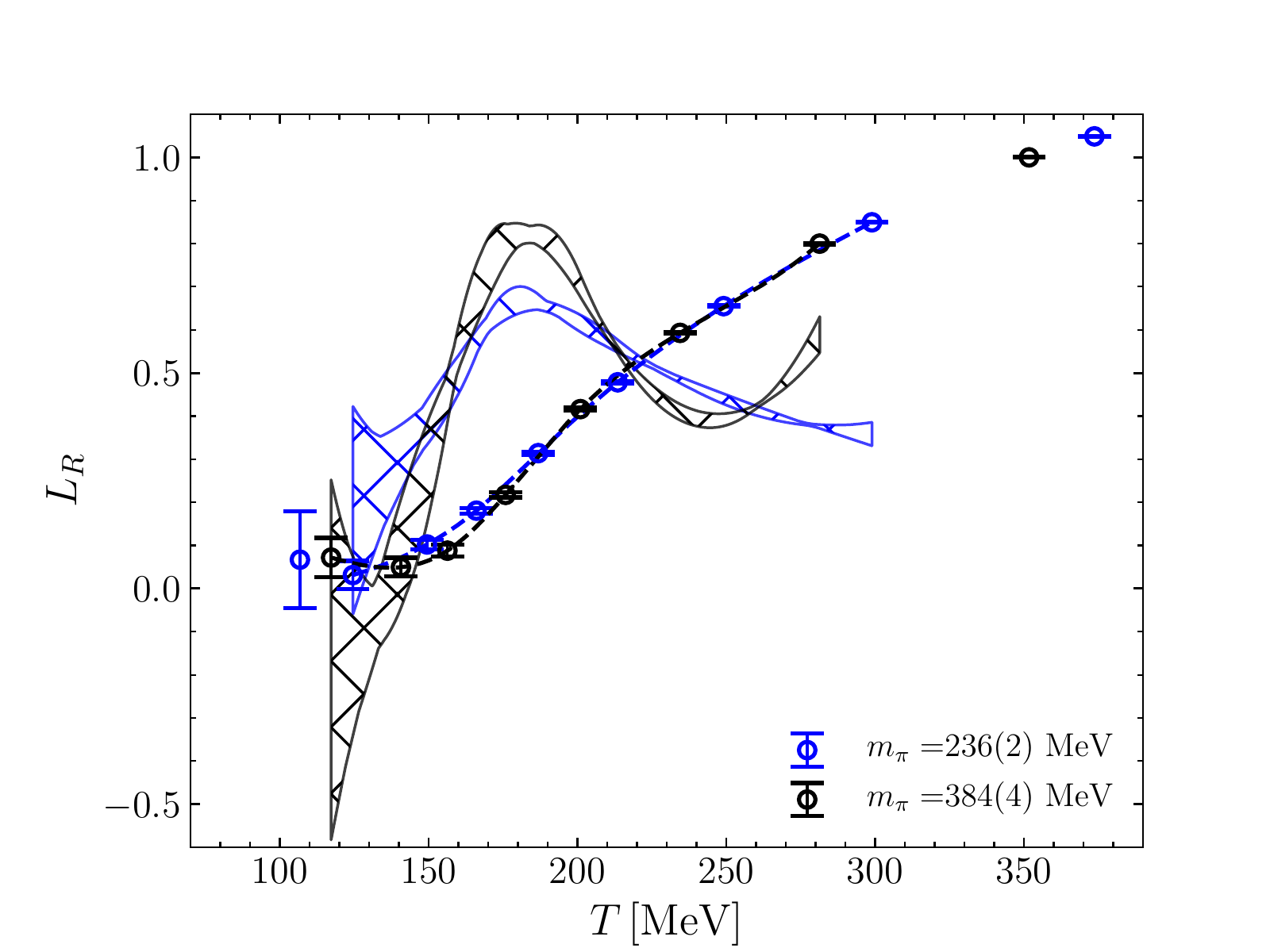}
\end{center}
  \caption{Renormalised Polyakov loop $L_R$ on the Gen2 (heavier pion) and Gen2L (lighter pion) ensembles. Data points are connected via cubic splines, excluding the points at the lowest and highest temperatures. The hashed regions indicate the uncertainties in its derivatives needed to locate an inflection point.}
    \label{fig:Polyakov_loop}
\end{figure}

The free energy contains an additive divergence~\cite{Borsanyi:2010bp,Borsanyi:2010cj,Bazavov:2014pvz}, which results in a multiplicative, temperature-dependent renormalisation of the Polyakov loop. Following the same procedure as in our earlier Gen2 analysis~\cite{Aarts:2014nba}, the renormalised Polyakov loop is defined as
\be 
\label{eq:L_R_def}
L_R = e^{-F^q_R/T} = e^{-(F^q+\Delta F^q)/T} = Z_L^{N_\tau} L_{\rm bare},
\ee
which relates $\Delta F^q$ to $Z_L$. In turn, $Z_L$ may be fixed by imposing a renormalisation condition at an (arbitrary) reference temperature $T_*$, 
\be
L_R(T_*) \equiv  \mbox{constant}.
\ee
Here we follow Ref.~\cite{Aarts:2014nba} (Fig. 1, scheme A) and fix $L_R(T_*)=1$ at $T_*=352$ MeV ($N_\tau^* = 16$) for Gen2. Since the temporal lattice spacing is different for Gen2L, the corresponding value of $N_\tau^*$ is no longer an integer. However, since any $T_*$ may be set as a reference point, this does not create a problem.

The renormalised Polyakov loop is shown for both Gen2 and Gen2L in Fig.~\ref{fig:Polyakov_loop}. At high temperature, the results for the two generations are in good agreement, emphasising the importance of the renormalisation. At lower temperatures, there is a slight difference, indicating a dependence on the pion mass in the crossover region. Fitting the data with cubic splines allows for an extraction of the inflection point, using the derivative of the spline.  
For Gen2 the pseudocritical temperature was estimated to be $\Tpc^P=185(4)\,$MeV \cite{Aarts:2014nba}, where the uncertainty reflected the spread between different renormalisation schemes but did not include statistical uncertainties. Here we determine the statistical uncertainty using a bootstrap analysis. Choosing Scheme A as above we find $\Tpc=183^{+5}_{-8}$ MeV for Gen2 and $183^{+6}_{-3}$ MeV for Gen2L, where the uncertainties are now purely statistical. This implies that the Polyakov loop is not sensitive to the pion mass in this regime. However, it should be noted that for this observable the transition region is rather broad, reflecting the fact that it is not an order parameter. This result of course provides an important motivation to focus on observables linked to chiral symmetry.

\begin{figure}[t]
\begin{center}
\includegraphics[width=0.48\textwidth]{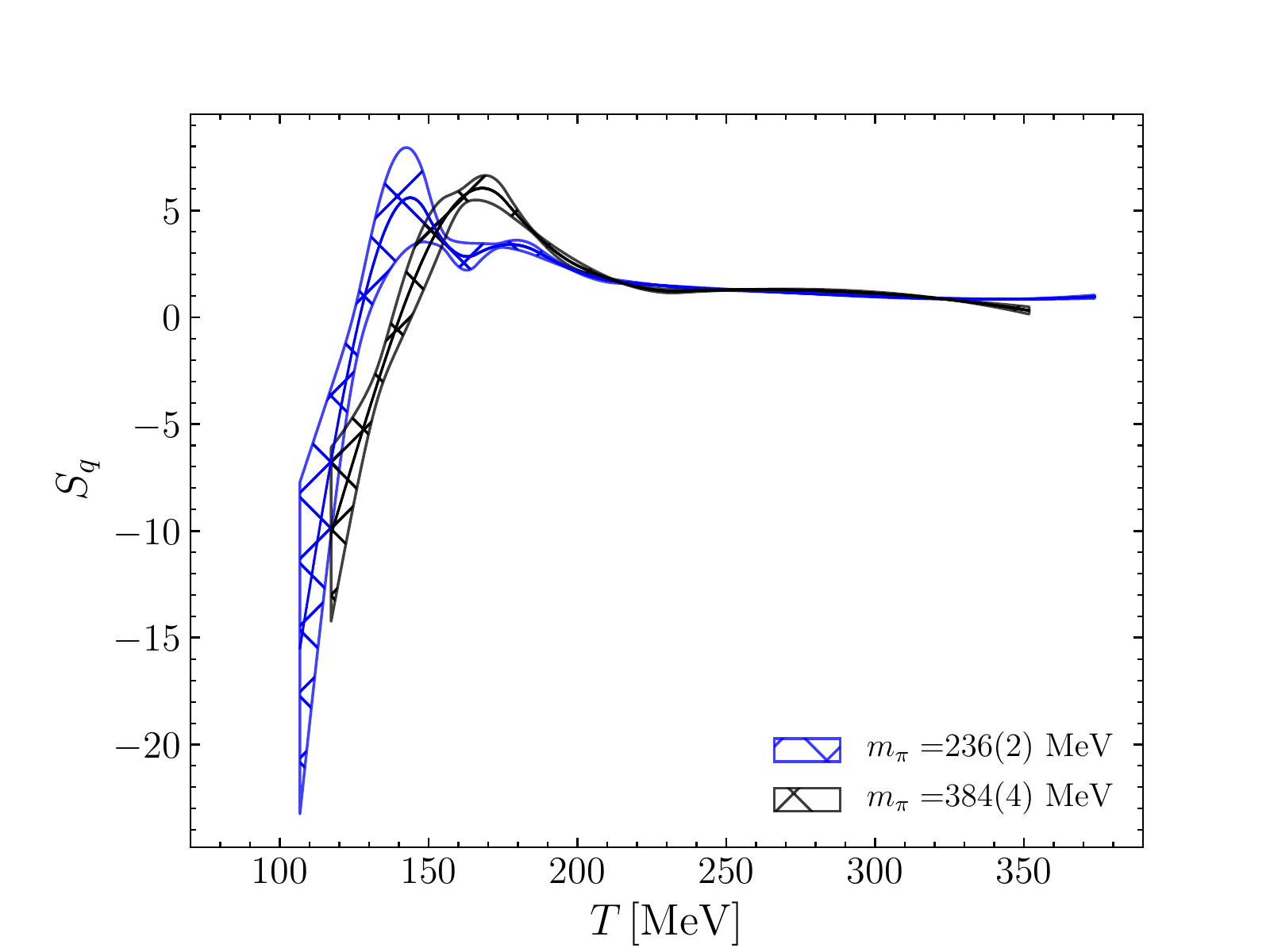}
\end{center}
  \caption{Single heavy-quark entropy $S_q$ on the Gen2 and 2L ensembles. The maxima are located at $\Tpc = 168(5)$ MeV and 144(8) MeV respectively. The hashed regions indicate the uncertainties.
  }
  \label{fig:entropy}
\end{figure}

Before doing so, however, we present one more result linked to the Polyakov loop, namely the entropy of a single, infinitely heavy quark. Following Refs.~\cite{Bazavov:2016uvm, Weber:2016fgn}, this entropy is defined as
\be
\label{eq:S_q_definition}
S_q = - \frac{\partial F^q_R}{\partial T}= \frac{\partial}{\partial T} \left( T\ln L_R \right),
\ee
and our results for $S_q$ are presented in Fig.~\ref{fig:entropy}. An estimate of the transition temperature is provided by its peak~\cite{Bazavov:2016uvm}. We find $\Tpc = 168(5)$ MeV for Gen2 and 144(8) MeV for Gen2L respectively. Hence in this case a clear pion mass dependence can be observed. The values for $\Tpc$ obtained in this section, along with those obtained below, are summarised in Table~\ref{tab:Tc_table} in Sec.~\ref{sec:summary}, where they will be compared in more detail.

\section{Susceptibilities}
\label{sec:sus}

To study the thermodynamic properties, we now discuss susceptibilities, {\it i.e.}, fluctuations of light and strange quark number, as well as of baryon number, electric charge and isospin. 
These are defined in the usual way (see, {\em e.g.}, Ref.\ \cite{Aarts:2014nba}) via the quark number density and quark number susceptibilities,
\begin{equation}
 \label{chi_ij}
 n_f = \frac{T}{V} \frac{\partial \ln{Z}}{\partial \mu_f},
 \quad\quad\
 \chi_{ff'} = \frac{T}{V} \frac{\partial^2 \ln{Z}}{\partial \mu_f\partial \mu_{f'} } =  \frac{\partial n_f}{\partial \mu_{f'}},
\end{equation} 
where $Z$ is the partition function, $V$ the spatial volume,  and $\mu_f$ the quark 
chemical potentials for flavours $f \in \{ u,d,s \}$. 
Note that baryon ($B$), isospin ($I$) and electrical charge ($Q$) chemical potentials are related to the quark chemical potentials as
\bea
\nn
&& \mu_u= \frac{1}{3}\mu_B+\frac{2e}{3}\mu_Q+\frac{1}{2}\mu_I, 
\qquad
\mu_s = \frac{1}{3}\mu_B-\frac{e}{3}\mu_Q, \\
&&\mu_d = \frac{1}{3}\mu_B-\frac{e}{3}\mu_Q-\frac{1}{2}\mu_I.
\eea
Here the electrical charge of the quark is denoted as $eq_f$, with $e$ the elementary charge and $q_f=2/3$ or  $-1/3$ its fractional charge.

\begin{figure}[t]
\begin{center}
\includegraphics[width=0.48\textwidth]{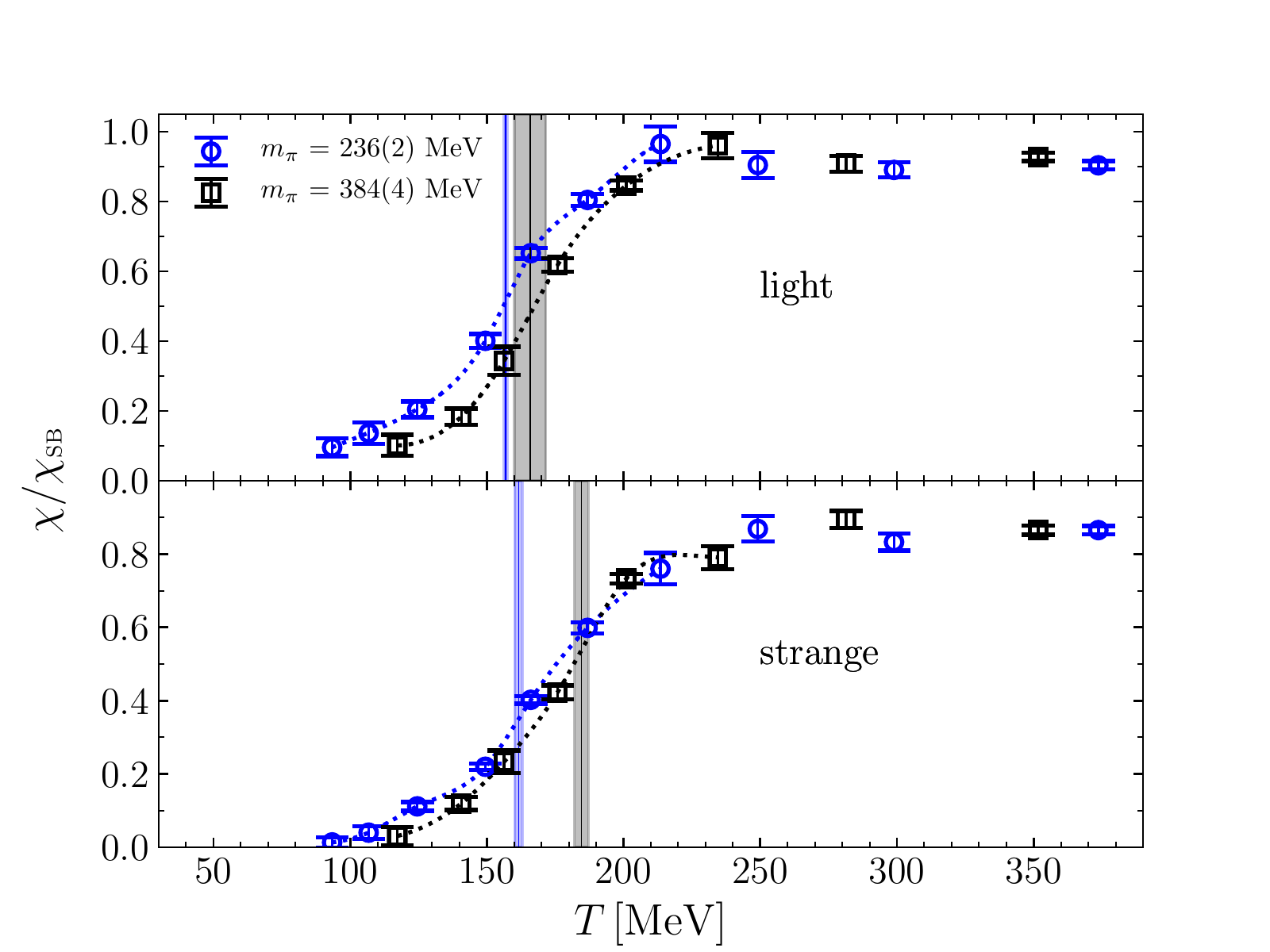}
\end{center}
     \caption{Comparison of light and strange quark number susceptibilities 
     for both sets of ensembles. The results are normalised with the respective quantities on the lattice for massless quarks in the Stefan--Boltzmann limit. Dotted lines represent interpolations by cubic splines. Vertical lines indicate the inflection point.
     }
    \label{fig:quarknumber_suscept}  
\end{figure}

Quark number susceptibilities for flavour $f$ are given by $\chi_{ff}$, while for baryon number, isospin and charge susceptibility, we find \cite{Aarts:2014nba}
\bea
\nn
&& \chi_B = \frac{1}{9}\sum_{f,f'} \chi_{ff'}, 
\qquad 
\chi_I = \frac{1}{4}\left( \chi_{uu} + \chi_{dd} - 2\chi_{ud} \right), \\
\qquad
&&\chi_Q =  e^2\sum_{f,f'} q_f q_{f'} \chi_{ff'}.
\eea
We follow the approach described in the previous study~\cite{Giudice:2013fza,Aarts:2014nba}, increasing the number of configurations and stochastic vectors for Gen2 substantially (see Table \ref{tab:Gens}) and extending the calculation to the new Gen2L ensembles.
Overall, the computational cost is dominated by the stochastic estimates of disconnected contributions \cite{Giudice:2013fza,Aarts:2014nba}. The only exception here is the isospin susceptibility $\chi_I$, where the disconnected parts cancel out in the case of degenerate light quarks. Stochastic estimators with Gaussian random vectors are employed in calculations; the number of vectors for each temperature may be found in  Table \ref{tab:Gens}. The signal-to-noise ratio for all susceptibilities allows us to interpolate and extract inflection points, with the baryon number susceptibility exhibiting the largest statistical fluctuations.

\begin{figure}[t]
\begin{center}
\includegraphics[width=0.48\textwidth]{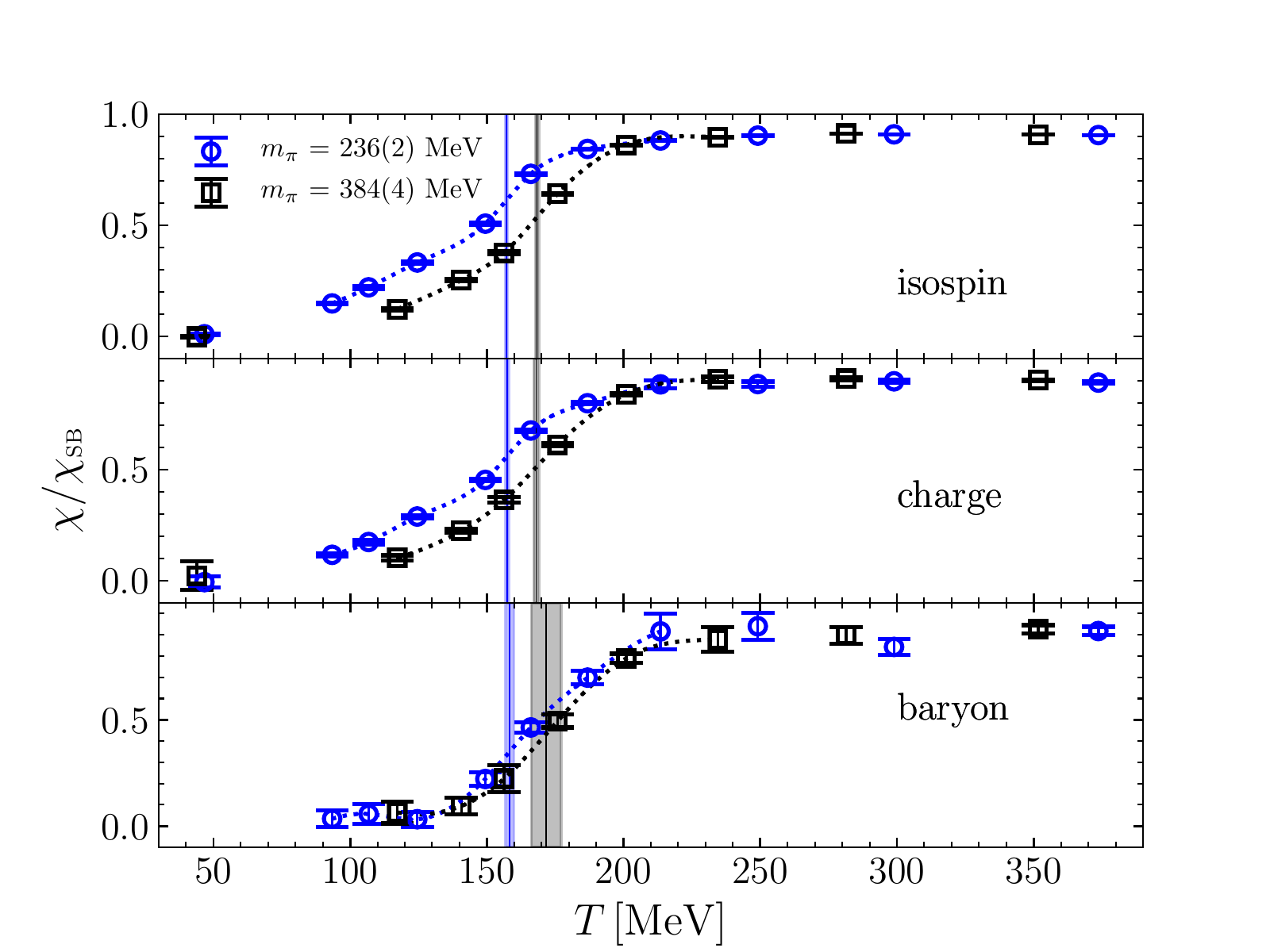}
\end{center}
     \caption{As in Fig.~\ref{fig:quarknumber_suscept}, for
     the isospin, charge and baryon number susceptibilities.
     }
    \label{fig:quarknumber_suscept2}  
\end{figure}

The results are presented in Figs.~\ref{fig:quarknumber_suscept} and \ref{fig:quarknumber_suscept2}, for the light and strange quark number susceptibilities and the isospin, charge and baryon number susceptibilities respectively. The susceptibilities are normalised with the corresponding quantities in the Stefan--Boltzmann limit for massless Wilson quarks on lattices with  the same geometry, using the renormalised anisotropy. The qualitative behaviour is the same for the heavier and the lighter pion masses; the main difference is the shift of the transition region to lower temperature. The effect of reducing the light quark mass is (marginally) the most pronounced for the isospin susceptibility. At high temperature, all susceptibilities approach the Stefan--Boltzmann limit from below. Again, the effect of reducing the light quark mass is most pronounced for the isospin susceptibility.

As a pragmatic definition for the transition temperature, we have fitted the data with cubic splines and extracted the inflection point. These temperatures are indicated with the vertical lines and are summarised in Table~\ref{tab:Tc_table} in Sec.~\ref{sec:summary}. Statistical errors are estimated via bootstrap. As expected, reducing the light quark masses brings the pseudocritical temperatures determined from the inflection points closer to the one observed for physical quark masses~\cite{Borsanyi:2010bp,Bazavov:2014pvz}. A more detailed discussion will be given in Sec.~\ref{sec:summary}.


\section{Chiral condensate and susceptibility}
\label{sec:chiral}

The key physical quantities used to study chiral properties of the system are the chiral condensate and its corresponding susceptibility,
\be \label{eq:ch_cond}
\langle \bar \psi_f \psi_f \rangle = \frac{T}{V} \frac{\partial \ln Z}{\partial m_f}\,
\qquad\quad 
\chi_{\bar\psi\psi} = \frac{T}{V} \frac{\partial^2 \ln Z}{\partial m_f^2}.
\ee
Both quantities contain additive and multiplicative divergences, which are regularised by the lattice cutoff. Since in the fixed-scale approach the lattice spacing is identical for all temperatures, a complete renormalisation is not required when we are only interested in extracting the pseudocritical temperature. However, for a more detailed comparison between the two generations --- with slightly different lattice spacings --- renormalisation  is necessary.

\begin{figure}[t]
\begin{center}
 \includegraphics[width=0.48\textwidth]{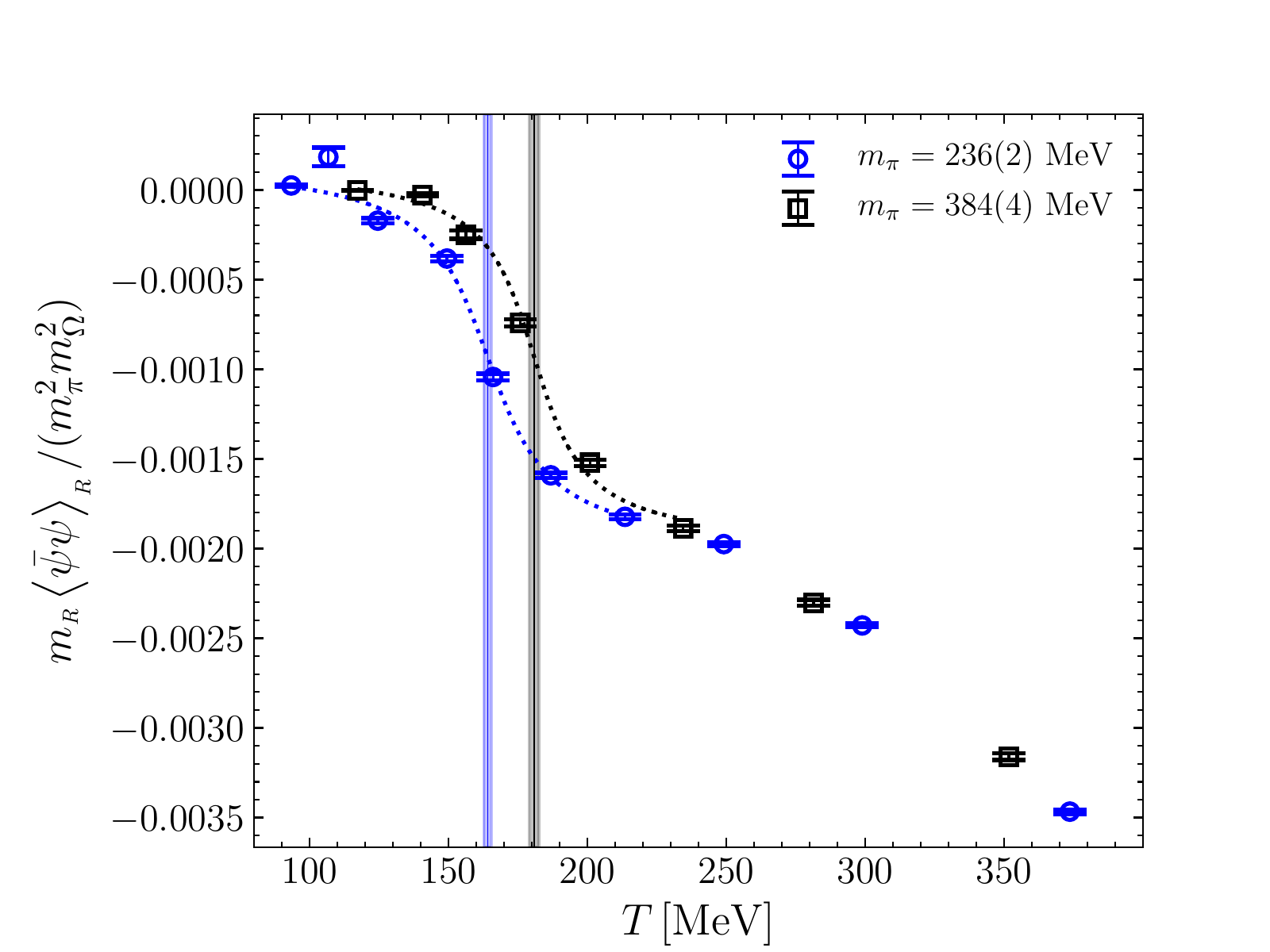}
\end{center}
   \caption{Renormalised chiral condensate  for $N_f=2$ light quarks in the combination $m_R \bra\bar\psi\psi\ket_R(T)/(m_\pi^2m_\Omega^2)$, for both sets of ensembles. 
   The dashed lines are fits according to Eq.~(\ref{eq:fitcc}), discarding the two/three highest points for Gen2/2L. Vertical lines indicate the inflection points.
   }
  \label{fig:ch_cond_light}
\end{figure}

To renormalise the chiral condensate we follow Ref.~\cite{Borsanyi:2012uq}, which builds on the formulation laid out in Ref.~\cite{Giusti:1998wy}. In this approach, additive divergences are cancelled by a zero-temperature subtraction, while multiplicative divergences are absorbed via the quark mass. Here we summarise the main equations. The subtracted chiral condensate is defined as
\be
\Delta_{\bar\psi\psi}(T) = \bra \bar \psi_l \psi_l \ket(T) - \bra \bar \psi_l \psi_l \ket(T=0),
\ee
where $\bra \bar \psi_l \psi_l \ket$ is the bare chiral condensate for $N_f=2$ degenerate light flavours, 
{\em i.e.}, 
\be \label{eq:ch_ll_definition}
\bra \bar \psi_l \psi_l \ket = \bra \bar \psi_u \psi_u \ket + \bra \bar \psi_d \psi_d \ket.
\ee
The subtracted pseudoscalar susceptibility is defined as
\bea
\nn
\Delta_{PP}(T) = && \int d^4x\, \bra P(x) P(0)\ket(T) \\
\label{eq:delta_PP}
&&-  \int d^4x\, \bra P(x) P(0)\ket(T=0),
\eea
where $P(x)$ is the bare pseudoscalar density 
\be
P(x) =  \frac{1}{N_f}\left( \bar \psi_u \gamma_5 \psi_u + \bar \psi_d \gamma_5 \psi_d\right).
\ee
Both quantities are related to the product of the renormalised quark mass $m_R$ and the renormalised subtracted chiral condensate $\bra\bar\psi\psi\ket_R(T)$, via~\cite{Giusti:1998wy,Borsanyi:2012uq}
\bea
&& m_R \bra\bar\psi\psi\ket_R(T) = 2N_f m^2_{PCAC} Z_A^2 \Delta_{PP}(T ),\\
&& m_R \bra\bar\psi\psi\ket_R(T) = m_{PCAC}Z_A\Delta_{\bar\psi\psi}(T) + \ldots
\eea
where $m_{PCAC}$ is the PCAC mass, $Z_A$ is a finite renormalisation constant, and the $\ldots$ vanish in the continuum limit.
Following Ref.~\cite{Borsanyi:2012uq}, the product of the renormalised mass and condensate can now be obtained from the ratio
\be
 \label{eq:mR_chcondR}
 m_R \bra\bar\psi\psi\ket_R(T) = \frac{\Delta_{\bar\psi\psi}^2(T) }{2 N_f \Delta_{PP}(T)} + \ldots,
 \ee
where the (bare) quantities on the RHS can be computed directly and there is no need to determine $m_{PCAC}$ and $Z_A$ separately. 

\begin{figure}[t]
\begin{center}
   \includegraphics[width=0.48\textwidth]{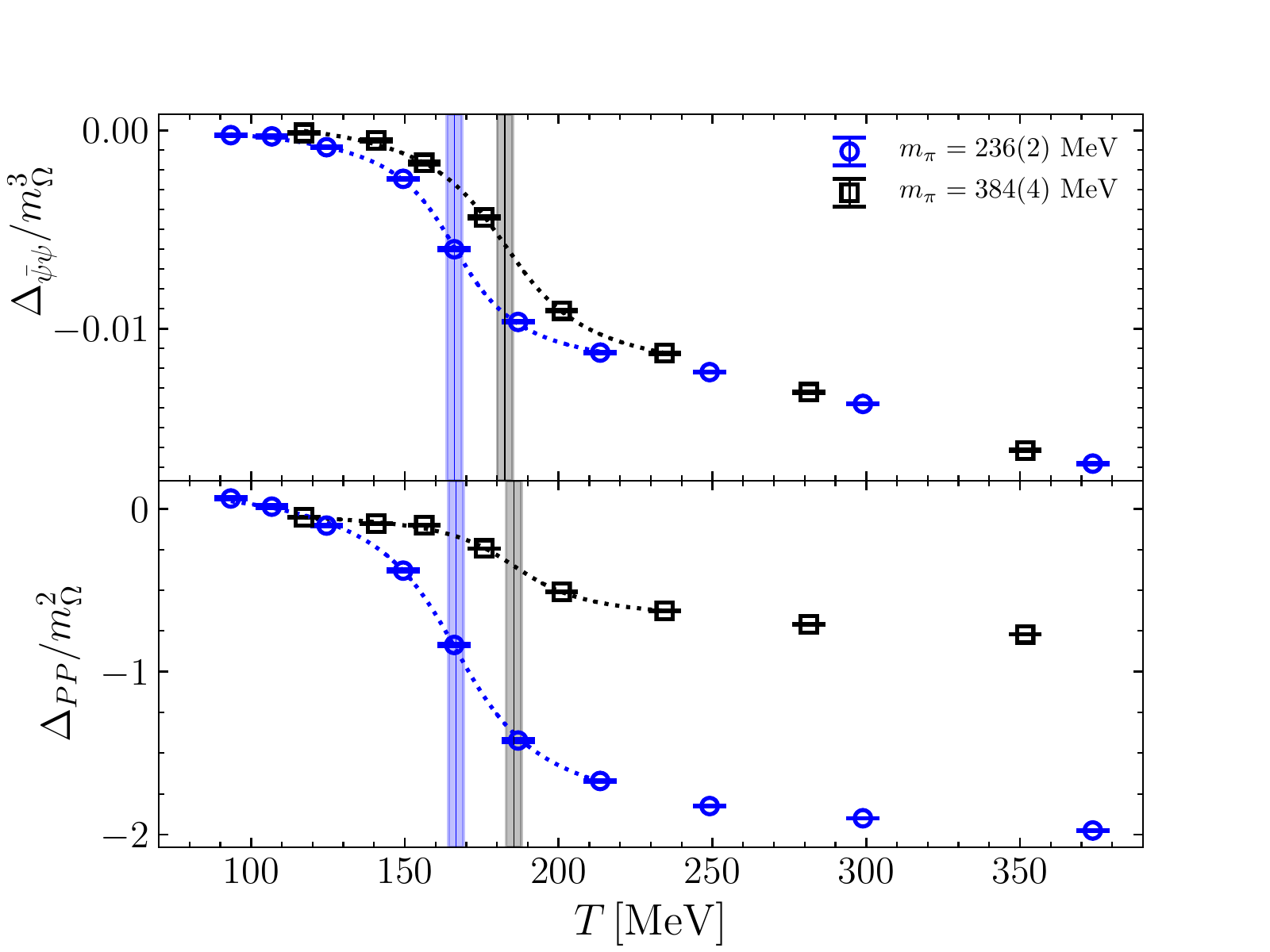}
\end{center}
   \caption{
   Dimensionless combinations $\Delta_{\bar\psi\psi}(T) / m_\Omega^3$ for the bare chiral condensate (above), see Eq.\,(\ref{eq:ch_ll_definition}), and $\Delta_{PP}(T) / m_\Omega^2$ for the pion susceptibility (below), see Eq.\,(\ref{eq:delta_PP}), for both sets of ensembles. Dashed and vertical lines are as in the preceding figure.
   }
  \label{fig:ch_cond_light2}
\end{figure}

The result is presented in Fig.~\ref{fig:ch_cond_light}. It is made dimensionless by dividing with $m_\pi^2 m_\Omega^2$, using the ``zero-temperature" values for each ensemble, such that the ratio is finite in the chiral limit. We note that the two sets of points agree with each other, except that the transition region is shifted to lower temperature for the lighter pion mass. To extract the pseudocritical temperature, we fit the data points to the Ansatz
\be
\label{eq:fitcc}
\frac{m_R \bra\bar\psi\psi\ket_R(T)}{m_\pi^2 m_\Omega^2} = c_0 + c_1\arctan\left[ c_2(T - \Tpc) \right],
\ee
discarding the two (three) highest temperature points for Gen2 (Gen2L). These fits, with $\chi^2/d.o.f. \approx 0.8$, yield $\Tpc^{\bar\psi\psi} = 181(2)$ MeV for Gen2 and $164(2)$ MeV for Gen2L (see Table~\ref{tab:Tc_table}). We will discuss these results further in Sec.~\ref{sec:summary}.

To further analyse the effect of renormalisation and verify that the pseudocritical temperature does not depend on the choice of observable in the fixed-scale approach, we show the bare subtracted chiral condensate $\Delta_{\bar\psi\psi}(T)$ and pseudoscalar susceptibility $\Delta_{PP}(T)$ separately in Fig.~\ref{fig:ch_cond_light2}. While the details of the data points now depend on the ensemble ({\em i.e.} the lattice spacing), the pseudocritical temperatures do not. Using again the fit (\ref{eq:fitcc}), we find, from $\Delta_{\bar\psi\psi}$,  $\Tpc = 183(3)$ MeV for Gen2 and 166(2) MeV for Gen2L, and from  $\Delta_{PP}(T)$, $\Tpc = 186(2)$ MeV for Gen2 and 166(2) MeV for Gen2L, which are consistent with the results given above,
as it should be.

\begin{figure}[t]
  \begin{center}
  \includegraphics[width=0.48\textwidth]{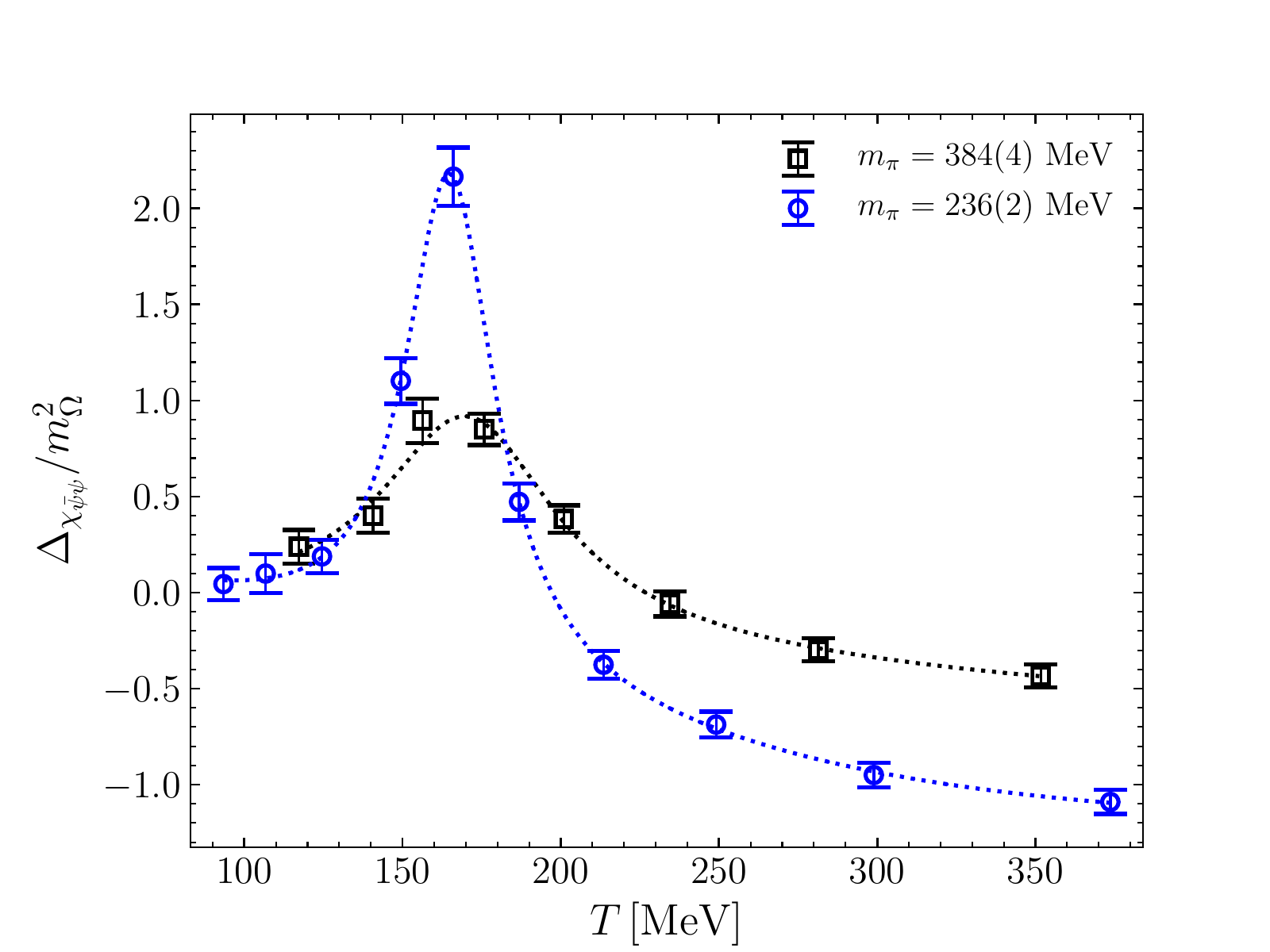}
  \end{center}
  \caption{Subtracted chiral susceptibility $\chi_{\bar\psi\psi}(T)-\chi_{\bar\psi\psi}(T=0)$ for light degenerate quarks ($N_f=2$), for both ensembles. It was normalised on the relevant mass on $\Omega$-baryon to make the quantity dimensionless. The dotted lines are fits according to Eq.~(\ref{eq:fitchicc}).   
 }
  \label{fig:ch_susceptibility_l}
\end{figure}

In Fig.~\ref{fig:ch_susceptibility_l} we present the chiral susceptibility for the two light flavours, with the value at ``zero temperature'' subtracted (using the $N_\tau = 128$ results for both generations), but without any multiplicative renormalisation. We note that we show the full susceptibility, {\em i.e.}, the sum of the connected and disconnected contributions (the former shows essentially no sensitivity to the crossover). The peak in the susceptibility is considerably more pronounced for the lighter pion. To extract the corresponding pseudocritical temperature, we used the fit
\bea
\nn
&& \chi_{\bar\psi\psi}(T) -\chi_{\bar\psi\psi}(0) = \\
&&  \frac{c_0}{c_1 + (T - \Tpc)^2}   
  + c_2 + c_3\tanh\left[c_4(T - \Tpc)\right].
\label{eq:fitchicc}
\eea
Taking $c_3=0$ yields an adequate fit in the temperature region near the peak, using 5 points for both Gen2 and Gen2L, giving $\Tpc^{\chi_{\bar\psi\psi}} = 170(3)$ MeV for Gen2 and 165(2) MeV for Gen2L (see Table~\ref{tab:Tc_table}). By adding the term proportional to $c_3$ a fit for all data points can be found. The difference in $\Tpc$ obtained using the first and the second form may be considered as an estimate of the systematic error. The fit with $c_3\neq 0$ provides a somewhat smaller $\chi^2/d.o.f.$ and increases the pseudocritical temperature by 2 MeV for both generations. We added this as an additional error for $\Tpc^{\chi_{\bar \psi \psi}}$ in  Table~\ref{tab:Tc_table}. One may observe that the difference between the pseudocritical temperatures from the chiral susceptibility in Gen2 and Gen2L is very small compared to other fermionic observables, which may be explained by the absence of a clear peak for the larger pion mass.

We note here that we also calculated the chiral condensate and susceptibility for the strange quark. Since it turned out to be much noisier than the light quark quantities, we do not present it here.


\section{Parity doubling for octet and decuplet baryons}
\label{sec:parity}

As the final probe of the thermal transition we consider here the emergence of parity doubling in baryonic correlators, which is a signal of chiral symmetry restoration. 
We construct the baryon $R$ parameter~\cite{Datta:2012fz,Aarts:2015mma,Aarts:2017rrl,Aarts:2018glk} from the positive- and negative-parity correlators $G_+(\tau)$ and $G_-(\tau) = -G_+(1/T - \tau)$, according to
\be
\label{eqn:R_value}
R = \frac{\sum_n R(\tau_n)/\sigma^2(\tau_n)}{\sum_n 1/\sigma^2(\tau_n)},
\ee
where  $\sigma(\tau_n)$ denotes the statistical error for $R(\tau_n)$, and $R(\tau_n)$ is defined as
\be
\label{eqn:R_tau}
R(\tau_n) = \frac{G_+(\tau_n) - G_+(1/T - \tau_n)}{G_+(\tau_n) + G_+(1/T - \tau_n)}.
\ee
The sum over the time slices $\tau_n$ in Eq.~(\ref{eqn:R_value}) includes  $4 \leq n < N_\tau/2$ at all temperatures, to suppress lattice artefacts at small  values of $\tau_n$. Since $R(1/T-\tau)=-R(\tau)$, only time slices with $n<N_\tau/2$ contribute independently. The physical reason to introduce this $R$ parameter is as follows (for a detailed discussion, see Ref.~\cite{Aarts:2017rrl}): when chiral symmetry is unbroken, positive- and negative-parity correlators are degenerate and $R=0$. On the other hand, if chiral symmetry is broken, and $G_\pm(\tau)$ are dominated by their respective ground states,  and the mass of the negative-parity partner is substantially larger than the positive-parity one, then $R\simeq 1$. Hence the expectation is that this parameter is close to one in the hadronic phase and close to zero at high temperature, with a transition in the crossover region. This is indeed the case;
Refs.~\cite{Aarts:2015mma,Aarts:2017rrl,Aarts:2018glk} contain a discussion in the context of the Gen2 ensembles.

\begin{figure}[t]
\begin{center}
\includegraphics[width=0.48\textwidth]{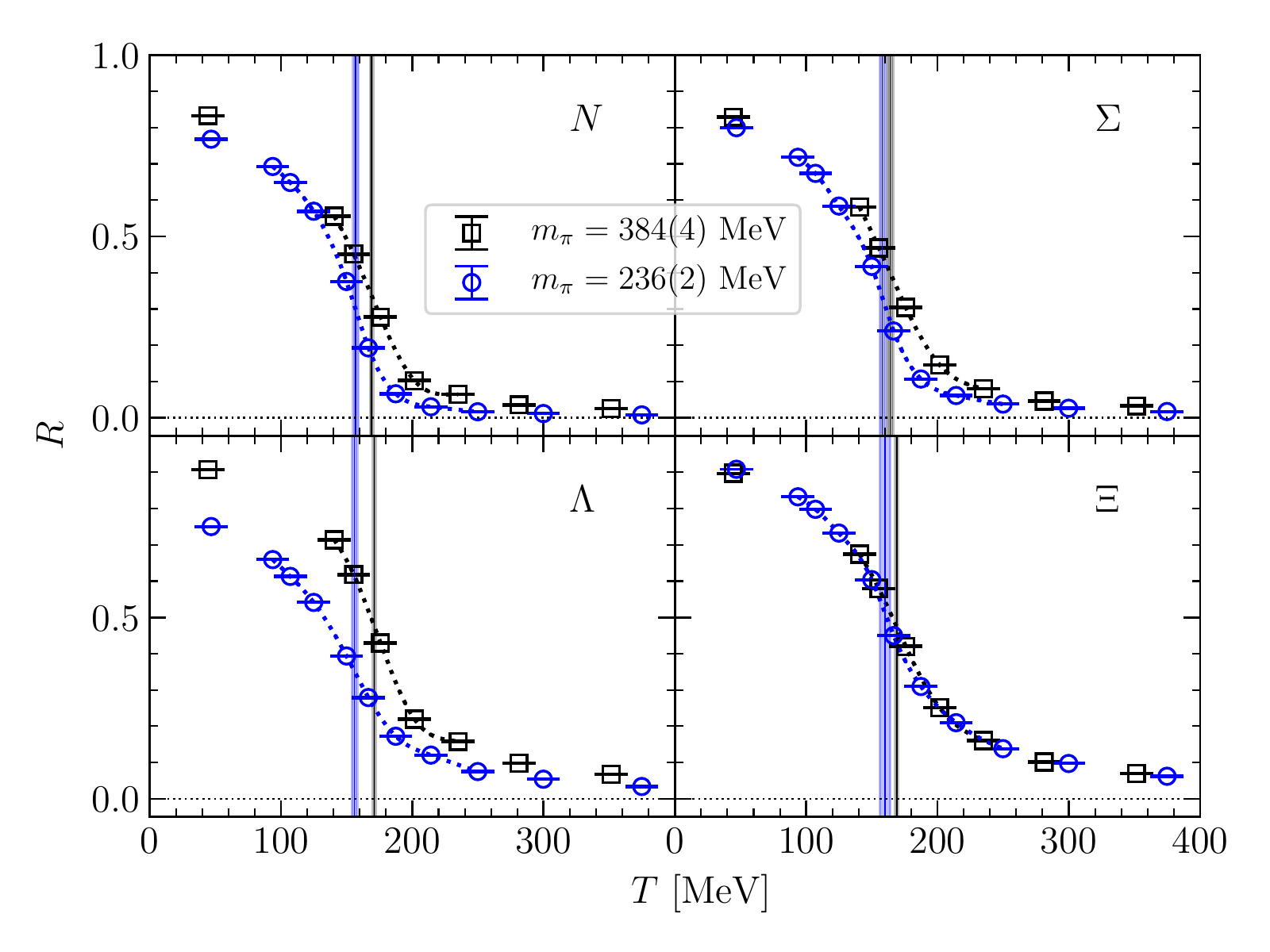}
\end{center}
    \caption{Parity-doubling $R$ parameter as a function of temperature for octet baryons, for  both sets of ensembles. Dotted lines represent interpolations by cubic splines. Vertical lines indicate the inflection point.}
    \label{fig:Roctet}
\end{figure}

\begin{figure}[t]
\begin{center}
\includegraphics[width=0.48\textwidth]{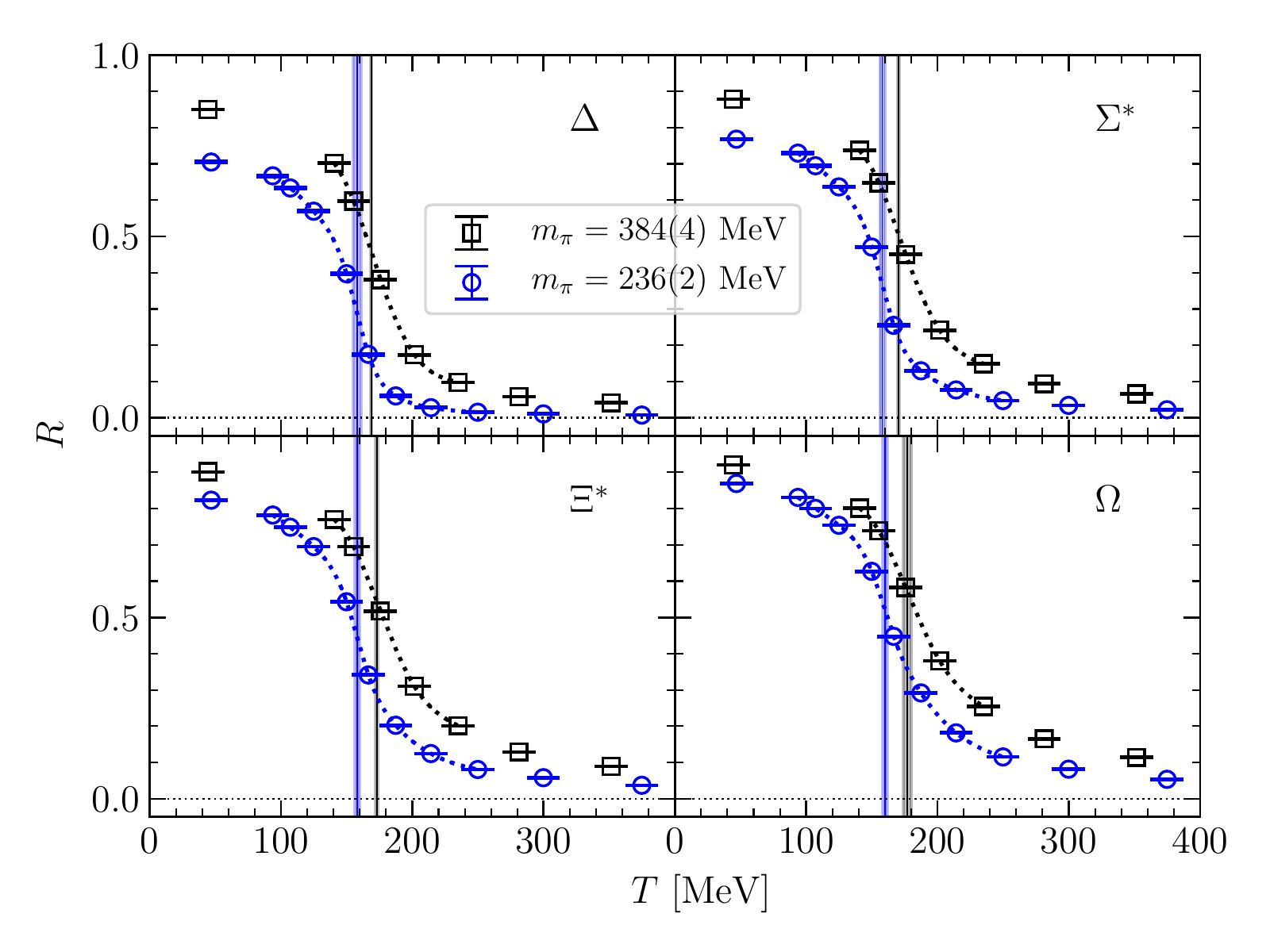}
\end{center}
    \caption{As in Fig.~\ref{fig:Roctet}, for decuplet baryons.
    }
    \label{fig:Rdecuplet}
\end{figure}

A comparison between both sets of ensembles is shown in Figs.~\ref{fig:Roctet} and \ref{fig:Rdecuplet}, for the octet and decuplet baryons respectively. The $R$ parameter is distinctly nonzero and close to one at the lowest temperature. Subsequently, as the temperature is increased, it goes towards zero in the quark-gluon plasma. With massive quarks, chiral symmetry is explicitly broken and $R\neq 0$ also in the high-temperature phase. This effect is expected to go away at very high temperature, as $m_q/T\to 0$. The consequence of the lighter quarks in Gen2L is visible especially in the nucleon and $\Delta$  channels, where $R$ approaches zero more rapidly. We note that the amount of smearing used to compute the baryon correlators has some effect on the detailed shape of the $R$ curve~\cite{Aarts:2015mma}. Here we are interested in the transition and the shift of the transition region towards lower temperature for the lighter pion. To analyse this, we have fitted the data with cubic splines and extracted the temperature of the inflection points, these are indicated with the vertical lines in Fig.~\ref{fig:Roctet} and Fig.~\ref{fig:Rdecuplet}, and are listed in the Table~\ref{tab:T_infl_R}. We also tried arctan-like fit of the form (\ref{eq:fitcc}), it lowers $\Tpc$ values by approximately 2 MeV compared to the ones from inflection, and the results remain the same within the errorbars. One may observe, that inflection point occurs at a lower temperature for the ensembles with the smaller pion mass. Moreover, the (weak) strangeness dependence observed for Gen2 in Ref.~\cite{Aarts:2018glk} is absent for Gen2L.

\begin{table}[b]
\begin{tabular}{|c|| cccc| }
\hline
 \; $T_{\rm inf}$[MeV] \;	 & $N$ & $\Sigma$ & $\Lambda$ & $\Xi$  	\\
 \hline
 Gen2 & 169(1) & 164(2) & 171(1) & 169(1)  	\\
 Gen2L &  157(2) & 158(2) & 156(2) & 160(4)   \\
\hline
 $T_{\rm inf}$[MeV]	 &  $\Delta$ & $\Sigma^*$ & $\Xi^*$	& $\Omega$ 	\\
 \hline
 Gen2 & \; 168.8(5) \; & \; 170.3(7) \; & \; 173(1)	 \;& \; 177(3) \;	 	\\
 Gen2L &   158(3) & 158(2) & 158(2) & 160(2) \\
\hline
\end{tabular}
\caption{Inflection-point temperatures $T_{\rm inf}$ of the $R$ parameter for the baryon channels considered, for both sets of ensembles.}
\label{tab:T_infl_R}
\end{table}

\section{Discussion and summary}
\label{sec:summary}

\begin{table}[b]
\begin{center}
    \begin{tabular}{| c|| c | c |  }
    \hline
                            & \multicolumn{2}{c|}{$T_{\rm pc}$ [MeV]}   \\
        \hline
         \; observable \;     & \; $m_\pi = 236(2)$ MeV \; & \; $m_\pi = 384(4)$ MeV \;  \\        \hline   
        $L_R$                    & $183^{+6}_{-3}$ & $183^{+5}_{-8}$ \\
        $S_q$                    & 144(8)      & 168(5) \\
        $\chi_{\rm light}$   	 & 157(1)      & 166(6) \\
        $\chi_{\rm strange}$     & 162(2)      & 184(3) \\
        $\chi_{\rm I}$       	 & 157.2(4)  & 168.4(6) \\
        $\chi_{\rm Q}$       	 & 157.5(6)  & 168.1(6) \\
        $\chi_{\rm B}$       	 & 158(2)      & 172(5) \\
        $\bra\bar\psi\psi\ket_R$ & 164(2)      & 181(2) \\
        $\chi_{\bar\psi\psi}$    & 165(2)(2)   & 170(3)(2) \\
        $R_{\rm baryon}$         & 156--160    & 164--177  \\
            \hline
    \end{tabular}
    \caption{Pseudocritical temperatures extracted from the renormalised Polyakov loop and the single heavy-quark entropy (see Sec.~\ref{sec:pol}), various susceptibilities (Sec.~\ref{sec:sus}), 
     the renormalised chiral condensate and its susceptibility (Sec.~\ref{sec:chiral}), 
     and the parity-doubling parameter $R$ for baryons (Sec.~\ref{sec:parity}). The error in the second brackets for the chiral susceptibility estimates the systematic uncertainty.}
    \label{tab:Tc_table}
\end{center}
\end{table}

\begin{figure}[t]
  \begin{center}
  \includegraphics[width=0.48\textwidth]{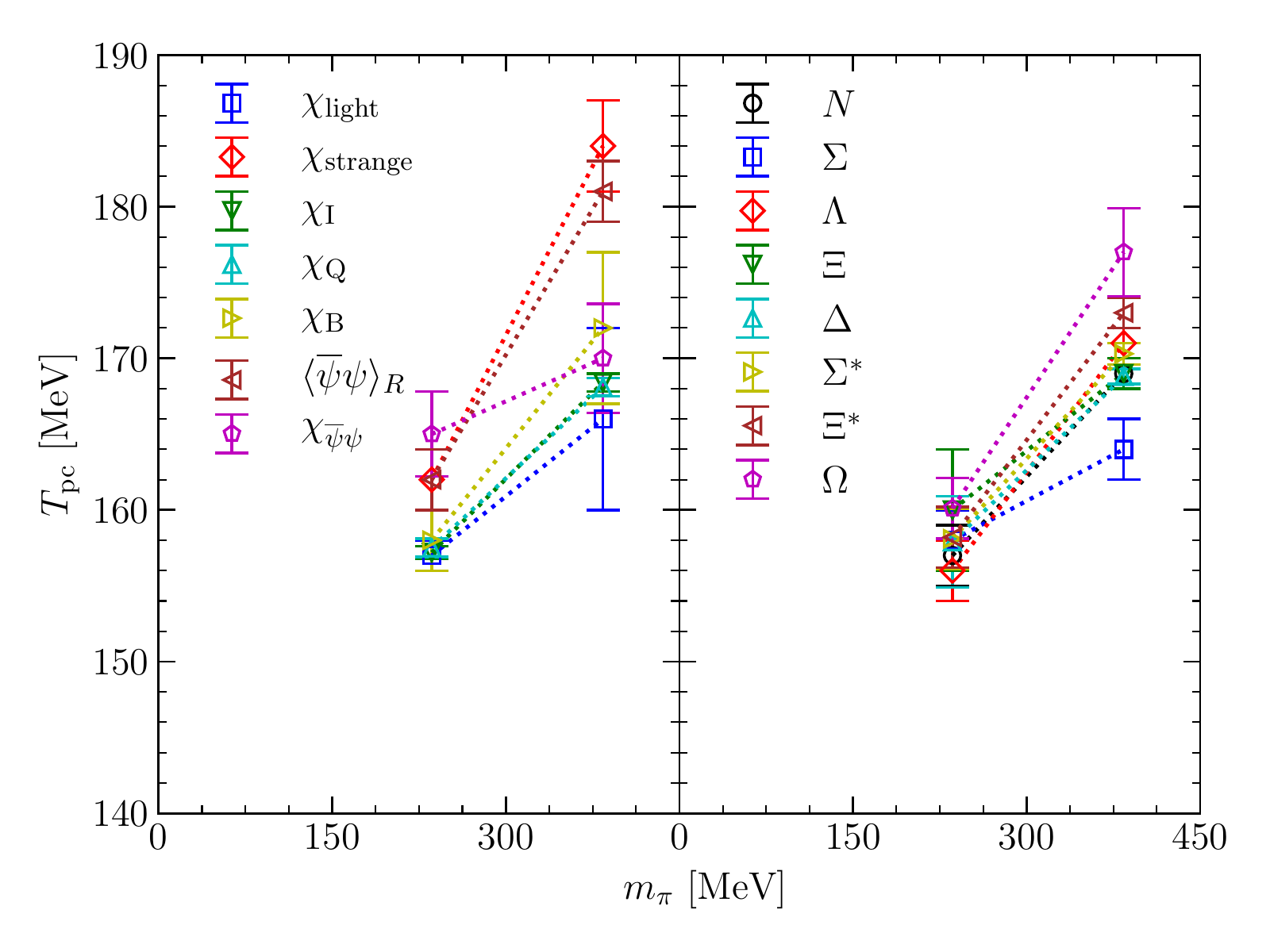}
  \caption{Estimates of the pseudocritical temperatures for the two pion masses considered, $m_\pi = 236(2),\,384(4)$ MeV, from different susceptibilities and the chiral condensate (on the left), and from the baryon $R$ parameter for different channels (on the right). Numerical values are summarised in the Tables~\ref{tab:T_infl_R} and \ref{tab:Tc_table}. Dotted lines are plotted to guide the eye.}
   \label{fig:Tpc}
  \end{center}
\end{figure}

We have analysed the thermal transition in QCD with $N_f=2+1$ flavours of improved Wilson fermions, using a wide range of observables related to the quark degrees of freedom, for two values of the pion mass. A summary of the pseudocritical temperatures found is provided in Table~\ref{tab:Tc_table}.
For the renormalised Polyakov loop we noted a smooth behaviour manifesting itself by an absence of a clearly distinguishable peak and a negligible dependence on the pion mass, see Fig.~\ref{fig:Polyakov_loop}. In contrast, the heavy-quark entropy shows a sharper crossover and a dependence on the pion mass, even though it is closely linked to the Polyakov loop. We continue the discussion by focusing on fermionic quantities, which are related to chiral symmetry. For these we observe that the temperature where the crossover occurs is reduced as the pion gets lighter, as expected, see Fig.~\ref{fig:Tpc}. Moreover, we note that the spread of the pseudocritical temperatures is reduced. This focusing of the pseudocricitical temperatures may be interpreted as a sign for the presence of a proper phase transition for very light quarks, {\it i.e.}, lighter than in nature. Indeed, it is expected that the chiral transition becomes either first order (ending in a second order point at a finite value of the pion mass) or second order (for a massless pion). We note here that the pseudocritical temperature extracted from the chiral susceptibility is somewhat of an outlier, taking on a smaller than expected value at the heavier pion mass --- note that from the theory of critical scaling, one expects $\Tpc^{\bar\psi\psi} < \Tpc^{\chi_{\bar\psi\psi}}$, which is not the case for Gen2.
This may be caused by the absence of a pronounced peak of the chiral susceptibility for the Gen2 ensembles, see Fig.~\ref{fig:ch_susceptibility_l}.

\begin{figure}[t]
  \begin{center}
  \includegraphics[width=0.48\textwidth]{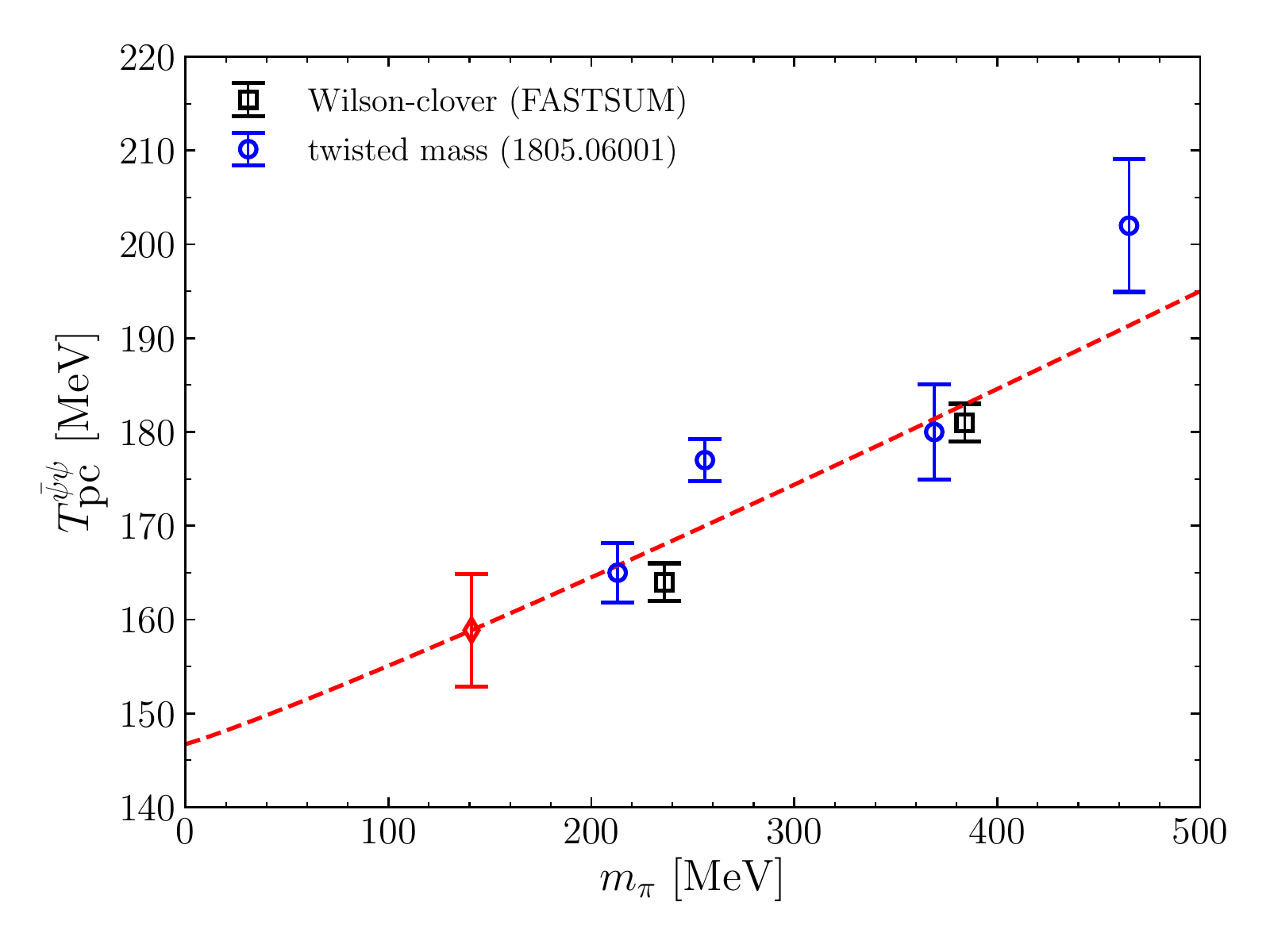}
  \caption{Estimates for the pion mass dependence of $\Tpc^{\bar\psi\psi}$, extracted from the inflection point of the renormalised chiral condensate, for $N_f=2+1+1$ twisted-mass~\cite{Burger:2018fvb} and $N_f = 2 + 1$ Wilson-clover (this work) fermions. The dashed line presents the fit (\ref{eq:fit}), with the diamond denoting the extrapolated value at the physical pion mass, $\Tpc^{\bar\psi\psi}=159(6)$ MeV.
}
  \label{fig:Tc_pion_mass}
  \end{center}
\end{figure}

With only two values of the pion mass, it is not possible to make a more quantitative statement about the critical temperature for either physical or massless quarks. However, to look for consistency we may compare our results with those obtained using other lattice fermion formulations (in particular of the Wilson type) in the same pion mass range. In Ref.~\cite{Burger:2018fvb}, the thermal transition was studied in $N_f=2+1+1$ QCD using twisted-mass fermions, for pions with masses between 213 and 466 MeV, at a single lattice spacing. In Fig.~\ref{fig:Tc_pion_mass}, we compare the pseudocritical temperatures extracted from the inflection point of the chiral condensate. We observe a consistent pion mass dependence, within the relatively large uncertainties. In an attempt to extrapolate to the physical point, we fit the data according to
\be
\label{eq:fit}
\Tpc^{\bar\psi\psi}(m_\pi) = T_0 + \kappa m_{\pi}^{2/\Delta},
\ee
where $\Delta = 1.833$ is fixed and represents the $O(4)$ universality class critical exponent~\cite{Engels:2011km}. The critical temperature in the chiral limit $T_0$ and the coefficient $\kappa$ are parameters to be determined. Fitting the six data points we find
\be
\kappa = 0.055(8)\,\mbox{MeV}^{1-2/\Delta},  
\qquad 
T_0 = 147(4)\,\mbox{MeV}.
\ee
Extrapolating this fit to the physical pion mass yields 
\be\label{eq:Tpc_phys_point}
\Tpc^{\bar\psi\psi} = 159(6) \,\mbox{MeV} \qquad\quad \mbox{(physical point)}\,, 
\ee
which is consistent with the results obtained from the chiral condensate by the Wuppertal-Budapest~\cite{Borsanyi:2010bp} and HotQCD~\cite{Bazavov:2018mes} collaborations, although we stress that no continuum extrapolation has been performed here. The error quoted in Eq.~(\ref{eq:Tpc_phys_point}) is statistical only.

As an outlook, we are in the process of studying the fate of hadrons at finite temperature on the Gen2L ensembles, with the lower pion mass, which is the main motivation for this work. Preliminary results for bottomonium have appeared in Ref.~\cite{Offler:2019eij}. In addition, we are currently tuning the  lattice parameters to simulate directly using physical quark masses, while still at fixed lattice spacing. This may also allow us to perform a proper investigation of critical scaling.

\section*{Acknowledgments}

We are grateful for support from STFC via grants ST/L000369/1 and ST/P00\-055X/1, the Swansea Academy for Advanced Computing, SNF, ICHEC, the European Research Council (ERC) under the European Union's Horizon 2020 research and innovation programme under grant agreement No 813942. AAN and MPL are grateful to COST Action CA15213 THOR and thank the Galileo Galilei Institute for Theoretical Physics for hospitality. SK is supported by the National Research Foundation of Korea under grant NRF-2018R1A2A2A05018231 funded by the Korean government (MEST). AAN also acknowledges the support by RFBR grant 18-32-20172 mol\_a\_ved. LKW is supported by the Key Laboratory of Ministry of Education of China under Grant No.~QLPL2018P01. 
We are grateful to DiRAC, HPC Wales, PRACE and Supercomputing Wales for the use of their computing resources. This work was performed using the PRACE Marconi-KNL resources hosted by CINECA, Italy and the DiRAC Extreme Scaling service and Blue Gene Q Shared Petaflop system at the University of Edinburgh operated by the Edinburgh Parallel Computing Centre. The DiRAC equipment is part of the UK’s National e-Infrastructure and was funded by by UK’s BIS National e-infrastructure capital grant ST/K000411/1, STFC capital grants ST/H008845/1 and ST/R00238X/1, and STFC DiRAC Operations grants ST/K005804/1, ST/K005790/1 and ST/R001006/1.


\appendix
\section{Lattice action and simulations}
\label{app:action}

Here we summarise the action formulation and its parameters, see also Refs.~\cite{Edwards:2008ja,Lin:2008pr,Aarts:2014nba}. The gauge action reads
\begin{eqnarray}
   S_G = \frac{\beta}{N_c \gamma_g}
    \sum_{x, i > \ipr} \left[
      \frac{c_0}{u_s^4} P_{i\ipr}(x) + \frac{c_1}{u_s^6} \left\{R_{i\ipr}(x) + R_{\ipr i}(x)\right\}
    \right] &&
    \nn \\
    +\frac{\beta\gamma_g}{N_c} \sum_{x,i} \left[
      \frac{c_0 + 4c_1}{u_s^2 u_{\tau}^2} P_{i4}
      + \frac{c_1}{u_s^4 u_{\tau}^2} \left\{R_{i4}(x) + R_{4i}(x)\right\} \right], &&
     \nn\\&&
     \label{eq:action_gauge} 
\end{eqnarray}
where $\beta=2N_c/g^2$ (with $N_c=3$) is the gauge coupling \footnote{The factor $1/N_c$ in front of the gauge action was missing in Ref.~\cite{Aarts:2014nba}.}, 
$\gamma_g$ is the bare gauge anisotropy, 
$u_s$ and $u_{\tau}$ are the tadpole improvement factors for the spatial and temporal links respectively, 
$c_{0,1}$ are the usual tree-level coefficients, and
$P_{\mu\nu}$ and $R_{\mu\nu}$ describe the 4-link plaquette and plain rectangular plaquette respectively,
\begin{eqnarray}
P_{\mu\nu} &=& N_c - \Tr\left[ U_{x, \mu} U_{x + \hmu, \nu} U^\dagger_{x + \hnu, \mu} U^\dagger_{x, \nu} \right], \\
R_{\mu\nu} &=& N_c \nn\\
&&- \Tr\left[ U_{x, \mu} U_{x + \hmu, \mu} U_{x + 2\hmu, \nu} U^\dagger_{x + \hnu + \hmu, \mu} U^\dagger_{x + \hnu, \mu} U^\dagger_{x, \nu} \right].
\nn
\end{eqnarray}
The indices $i,\, \ipr$ denote spatial directions ($i,\,\ipr = 1,2,3$) and the index $4$ denotes temporal direction in Eq.~(\ref{eq:action_gauge}) and below. Note that only plain rectangular plaquettes $R_{\mu\nu}$ were used, ``chair-like'' 6-link plaquettes were not included. The choice of parameter values used here is listed in Table~\ref{tab:parameters}.

Concerning the fermionic action, $S_F = \sum_{xy}\overline{\psi}_x D_{xy} \psi_y$, the Dirac operator reads
\bea
D &=& \hat{m}_0 + D_{W,4} + \frac{1}{\gamma_f} \sum_{i} D_{W,i} \nn\\
&&
  - \frac{c_\tau}{2} \sum_{i}  \sigma_{4i} \hat{F}_{4i}
  - \frac{c_s}{2\gamma_g} \sum_{i < \ipr} \sigma_{i\ipr} \hat{F}_{i\ipr},
 \label{eq:action_fermionic} 
 \eea
with 
\begin{eqnarray}
D_{W,4} &=& \half\left( 1 - \gamma_4 \right) U_{x,4} \delta_{x + \hhtime, y} +  \half \left( 1 + \gamma_4 \right) U^\dagger_{y,4} \delta_{x - \hhtime, y}, \nn \\
D_{W,i} &=&  \half\left( 1 - \gamma_i \right) U^{(2)}_{x,i} \delta_{x + \hat\imath, y} +  \half\left( 1 + \gamma_i \right) U^{(2)\dagger}_{y,i} \delta_{x - \hat\imath, y}, \label{eq:Dw_spatial}
\nn
\end{eqnarray}
and $\sigma_{\mu\nu} = i\left[ \gamma_{\mu},\,\gamma_{\nu} \right]/2$. Here $\hat{m}_0 = a_\tau m_f$ defines the bare quark mass, $\gamma_f$ sets the bare fermion anisotropy, and $D_{W,4}$ and  $D_{W,i}$ are the temporal and spatial Wilson terms respectively. It is important to note that these Wilson terms contain no tadpole improvement. The spatial links are stout smeared~\cite{Morningstar:2003gk} with two steps of smearing, using the weight $\rho = 0.14$, which is reflected as a superscript $U^{(2)}_{x,i}$ in (\ref{eq:Dw_spatial}). The Dirac operator (\ref{eq:action_fermionic}) also contains the clover terms $\hat{F}_{\mu\nu}$~\cite{Luscher:1996sc} consisting of four ``clover-like'' link paths,
\begin{eqnarray}
\hat{F}_{\mu\nu}(x) & = & \frac{i}{8} \underset{p = 1}{\overset{4}{\sum}} \Bigl[ U_{\mu\nu}^{(p)}(x) - U_{\mu\nu}^{(p)\dagger}(x) \Bigr], \\ \label{eq:clover_term_definition}
U_{\mu\nu}^{(1)}(x) & = & U_{x, \mu} U_{x+\hmu, \nu} U_{x+\hnu, \mu}^{\dagger} U_{x, \nu}^{\dagger}, \nn \\
U_{\mu\nu}^{(2)}(x) & = & U_{x, \nu} U_{x-\hmu+\hnu,\mu}^{\dagger} U_{x-\hmu, \nu}^{\dagger} U_{x-\hmu, \mu}, \nn \\
U_{\mu\nu}^{(3)}(x) & = & U_{x-\hmu, \mu}^{\dagger} U_{x-\hmu-\hnu, \nu}^{\dagger} U_{x-\hmu-\hnu, \mu} U_{x-\hnu, \nu}, \nn \\
U_{\mu\nu}^{(4)}(x) & = & U_{x-\hnu, \nu}^{\dagger} U_{x-\hnu, \mu} U_{x+\hmu-\hnu, \nu} U_{x, \mu}^{\dagger}. \nn
\end{eqnarray}
Note that the spatial links in the clover term are stout smeared in the same way as in $D_{W,i}$.
The factors $c_\tau$ and $c_s$ in front of the clover terms in Eq.~(\ref{eq:action_fermionic}) are the temporal and spatial clover coefficients respectively. They may be expressed as
\be
c_\tau = \half\left(\frac{\gamma_g}{\gamma_f}+\frac{1}{\xi_{\rm target}} \right)\frac{1}{\tilde u_s^2\tilde u_\tau},
\qquad
c_s = \frac{\gamma_g}{\gamma_f}\frac{1}{\tilde u_s^3},
\ee
where $\tilde u_{s,\tau}$ are the tadpole factors obtained with smeared links (see Table~\ref{tab:parameters}) and $\xi_{\rm target} = 3.5$ is the target anisotropy. The renormalised values of the anisotropy may be found in Table~\ref{tab:lattice_spacings}.

Finally we note that in the case of anisotropic lattices the bare quark mass is related to the hopping parameter $\kappa$ as follows~\cite{Chen:2000ej}:
\begin{equation}
\label{eq:kappa_m0_relation}
\frac{1}{2\kappa} = \hat{m}_0 + 1 + \frac{3}{\gamma_f},
\end{equation}
where $\gamma_f$ is again the bare fermion anisotropy. The choice of parameter values for the fermionic action can be found in the Table~\ref{tab:parameters}. Actually, the only difference between Gen2 and Gen2L action setup is the light quark mass $\hat{m}_{0, \rm light}$ (or, alternatively, $\kappa_{\rm light}$, using Eq.\ (\ref{eq:kappa_m0_relation}) to limited accuracy), because all other parameters including $\gamma_f$ remain the same.

\begin{table*}[t]
\centering
\begin{tabular}{| l | l |}
\hline
gauge coupling (fixed-scale approach)     & $\beta = 1.5$ \\
tree-level  coefficients 					& $c_0=5/3,\,c_1=-1/12$ \\
bare gauge, fermion anisotropy 			& $\gamma_g = 4.3$, $\gamma_f = 3.399$ \\
ratio of bare anisotropies          &  $\nu = \gamma_g / \gamma_f = 1.265$ \\
spatial tadpole (without, with smeared links) & $u_s = 0.733566$, $\tilde{u}_s = 0.92674$ \;  \\
temporal tadpole (without, with smeared links) \; & $u_\tau = 1$, $\tilde u_\tau = 1$  \\
spatial, temporal clover coefficient 			& $c_s = 1.5893$, $c_\tau = 0.90278$ \\
stout smearing for spatial links 			& $\rho = 0.14$, isotropic, 2 steps \\
bare light quark mass (Gen2, Gen2L) 		& $\hat m_{0, \rm light} = -0.0840,\,-0.0860$ \\
bare strange quark mass          		& $\hat m_{0, \rm strange} = -0.0743$ \\
light quark hopping parameter (Gen2, Gen2L) &  $\kappa_{\rm light} = 0.2780,\,0.27831$ \\
strange quark hopping parameter              &  $\kappa_{\rm strange} = 0.2765$ \\
\hline
\end{tabular}
\caption{Parameters in the lattice action (\ref{eq:action_gauge}) -- (\ref{eq:Dw_spatial}). Note that the bare fermion anisotropy is obtained as $\gamma_f=\gamma_g/\nu$.}
\label{tab:parameters}
\end{table*}

The Generation 2 ensembles were generated with the Chroma software~\cite{Edwards:2004sx}. 
To generate the Generation 2L ensembles with the lighter quarks, we have adapted openQCD~\cite{openqcd} code -- which at the time had more advanced algorithms for LA solvers compared to Chroma -- to include anisotropic lattices and stout-smeared gauge links. 
In addition, our fork makes use of AVX-512 optimisations, further improving performance on recent Intel Skylake and Knights Landing CPUs~\cite{Bennett:2018oyb}, which are deployed at DiRAC Extreme Scaling machines. This adaptation of openQCD is publicly available~\cite{fastsum,openqcd-fastsum}; it is an order of magnitude faster than the version of Chroma we employed in the past. Moreover, we introduced new modules to openQCD code, {\it e.g.} a stand-alone measurement code which constructs hadronic two-point functions~\cite{openqcd-hadspec} and allows to perform the calculations of correlation functions for various operators, with and without Gaussian smearing at the sources (sinks), using the definitions of Ref.~\cite{Leinweber:2004it}. This measurement code and other modules are available at the same location as our openQCD fork~\cite{fastsum}.

\bibliography{references}

\begin{thebibliography}{53}%
\makeatletter
\providecommand \@ifxundefined [1]{%
 \@ifx{#1\undefined}
}%
\providecommand \@ifnum [1]{%
 \ifnum #1\expandafter \@firstoftwo
 \else \expandafter \@secondoftwo
 \fi
}%
\providecommand \@ifx [1]{%
 \ifx #1\expandafter \@firstoftwo
 \else \expandafter \@secondoftwo
 \fi
}%
\providecommand \natexlab [1]{#1}%
\providecommand \enquote  [1]{``#1''}%
\providecommand \bibnamefont  [1]{#1}%
\providecommand \bibfnamefont [1]{#1}%
\providecommand \citenamefont [1]{#1}%
\providecommand \href@noop [0]{\@secondoftwo}%
\providecommand \href [0]{\begingroup \@sanitize@url \@href}%
\providecommand \@href[1]{\@@startlink{#1}\@@href}%
\providecommand \@@href[1]{\endgroup#1\@@endlink}%
\providecommand \@sanitize@url [0]{\catcode `\\12\catcode `\$12\catcode
  `\&12\catcode `\#12\catcode `\^12\catcode `\_12\catcode `\%12\relax}%
\providecommand \@@startlink[1]{}%
\providecommand \@@endlink[0]{}%
\providecommand \url  [0]{\begingroup\@sanitize@url \@url }%
\providecommand \@url [1]{\endgroup\@href {#1}{\urlprefix }}%
\providecommand \urlprefix  [0]{URL }%
\providecommand \Eprint [0]{\href }%
\providecommand \doibase [0]{http://dx.doi.org/}%
\providecommand \selectlanguage [0]{\@gobble}%
\providecommand \bibinfo  [0]{\@secondoftwo}%
\providecommand \bibfield  [0]{\@secondoftwo}%
\providecommand \translation [1]{[#1]}%
\providecommand \BibitemOpen [0]{}%
\providecommand \bibitemStop [0]{}%
\providecommand \bibitemNoStop [0]{.\EOS\space}%
\providecommand \EOS [0]{\spacefactor3000\relax}%
\providecommand \BibitemShut  [1]{\csname bibitem#1\endcsname}%
\let\auto@bib@innerbib\@empty
\bibitem [{\citenamefont {Aoki}\ \emph {et~al.}(2006)\citenamefont {Aoki},
  \citenamefont {Endr{\H o}di}, \citenamefont {Fodor}, \citenamefont {Katz},\
  and\ \citenamefont {Szabo}}]{Aoki:2006we}%
  \BibitemOpen
  \bibfield  {author} {\bibinfo {author} {\bibfnamefont {Y.}~\bibnamefont
  {Aoki}}, \bibinfo {author} {\bibfnamefont {G.}~\bibnamefont {Endr{\H o}di}},
  \bibinfo {author} {\bibfnamefont {Z.}~\bibnamefont {Fodor}}, \bibinfo
  {author} {\bibfnamefont {S.}~\bibnamefont {Katz}}, \ and\ \bibinfo {author}
  {\bibfnamefont {K.}~\bibnamefont {Szabo}},\ }\href {\doibase
  10.1038/nature05120} {\bibfield  {journal} {\bibinfo  {journal} {Nature}\
  }\textbf {\bibinfo {volume} {443}},\ \bibinfo {pages} {675} (\bibinfo {year}
  {2006})},\ \Eprint {http://arxiv.org/abs/hep-lat/0611014}
  {arXiv:hep-lat/0611014} \BibitemShut {NoStop}%
\bibitem [{\citenamefont {Bors\'anyi}\ \emph
  {et~al.}(2010{\natexlab{a}})\citenamefont {Bors\'anyi}, \citenamefont
  {Fodor}, \citenamefont {Hoelbling}, \citenamefont {Katz}, \citenamefont
  {Krieg}, \citenamefont {Ratti},\ and\ \citenamefont
  {Szabo}}]{Borsanyi:2010bp}%
  \BibitemOpen
  \bibfield  {author} {\bibinfo {author} {\bibfnamefont {S.}~\bibnamefont
  {Bors\'anyi}}, \bibinfo {author} {\bibfnamefont {Z.}~\bibnamefont {Fodor}},
  \bibinfo {author} {\bibfnamefont {C.}~\bibnamefont {Hoelbling}}, \bibinfo
  {author} {\bibfnamefont {S.~D.}\ \bibnamefont {Katz}}, \bibinfo {author}
  {\bibfnamefont {S.}~\bibnamefont {Krieg}}, \bibinfo {author} {\bibfnamefont
  {C.}~\bibnamefont {Ratti}}, \ and\ \bibinfo {author} {\bibfnamefont {K.~K.}\
  \bibnamefont {Szabo}} (\bibinfo {collaboration} {Wuppertal-Budapest}),\
  }\href {\doibase 10.1007/JHEP09(2010)073} {\bibfield  {journal} {\bibinfo
  {journal} {JHEP}\ }\textbf {\bibinfo {volume} {09}},\ \bibinfo {pages} {073}
  (\bibinfo {year} {2010}{\natexlab{a}})},\ \Eprint
  {http://arxiv.org/abs/1005.3508} {arXiv:1005.3508 [hep-lat]} \BibitemShut
  {NoStop}%
\bibitem [{\citenamefont {Bors\'anyi}\ \emph
  {et~al.}(2010{\natexlab{b}})\citenamefont {Bors\'anyi}, \citenamefont
  {Endr{\H o}di}, \citenamefont {Fodor}, \citenamefont {Jakovac}, \citenamefont
  {Katz}, \citenamefont {Krieg}, \citenamefont {Ratti},\ and\ \citenamefont
  {Szabo}}]{Borsanyi:2010cj}%
  \BibitemOpen
  \bibfield  {author} {\bibinfo {author} {\bibfnamefont {S.}~\bibnamefont
  {Bors\'anyi}}, \bibinfo {author} {\bibfnamefont {G.}~\bibnamefont {Endr{\H
  o}di}}, \bibinfo {author} {\bibfnamefont {Z.}~\bibnamefont {Fodor}}, \bibinfo
  {author} {\bibfnamefont {A.}~\bibnamefont {Jakovac}}, \bibinfo {author}
  {\bibfnamefont {S.~D.}\ \bibnamefont {Katz}}, \bibinfo {author}
  {\bibfnamefont {S.}~\bibnamefont {Krieg}}, \bibinfo {author} {\bibfnamefont
  {C.}~\bibnamefont {Ratti}}, \ and\ \bibinfo {author} {\bibfnamefont {K.~K.}\
  \bibnamefont {Szabo}},\ }\href {\doibase 10.1007/JHEP11(2010)077} {\bibfield
  {journal} {\bibinfo  {journal} {JHEP}\ }\textbf {\bibinfo {volume} {11}},\
  \bibinfo {pages} {077} (\bibinfo {year} {2010}{\natexlab{b}})},\ \Eprint
  {http://arxiv.org/abs/1007.2580} {arXiv:1007.2580 [hep-lat]} \BibitemShut
  {NoStop}%
\bibitem [{\citenamefont {Bazavov}\ \emph {et~al.}(2014)\citenamefont {Bazavov}
  \emph {et~al.}}]{Bazavov:2014pvz}%
  \BibitemOpen
  \bibfield  {author} {\bibinfo {author} {\bibfnamefont {A.}~\bibnamefont
  {Bazavov}} \emph {et~al.} (\bibinfo {collaboration} {HotQCD}),\ }\href
  {\doibase 10.1103/PhysRevD.90.094503} {\bibfield  {journal} {\bibinfo
  {journal} {Phys. Rev.}\ }\textbf {\bibinfo {volume} {D90}},\ \bibinfo {pages}
  {094503} (\bibinfo {year} {2014})},\ \Eprint {http://arxiv.org/abs/1407.6387}
  {arXiv:1407.6387 [hep-lat]} \BibitemShut {NoStop}%
\bibitem [{\citenamefont {Kogut}\ \emph {et~al.}(1982)\citenamefont {Kogut},
  \citenamefont {Stone}, \citenamefont {Wyld}, \citenamefont {Shigemitsu},
  \citenamefont {Shenker},\ and\ \citenamefont {Sinclair}}]{Kogut:1982fn}%
  \BibitemOpen
  \bibfield  {author} {\bibinfo {author} {\bibfnamefont {J.~B.}\ \bibnamefont
  {Kogut}}, \bibinfo {author} {\bibfnamefont {M.}~\bibnamefont {Stone}},
  \bibinfo {author} {\bibfnamefont {H.}~\bibnamefont {Wyld}}, \bibinfo {author}
  {\bibfnamefont {J.}~\bibnamefont {Shigemitsu}}, \bibinfo {author}
  {\bibfnamefont {S.}~\bibnamefont {Shenker}}, \ and\ \bibinfo {author}
  {\bibfnamefont {D.}~\bibnamefont {Sinclair}},\ }\href {\doibase
  10.1103/PhysRevLett.48.1140} {\bibfield  {journal} {\bibinfo  {journal}
  {Phys. Rev. Lett.}\ }\textbf {\bibinfo {volume} {48}},\ \bibinfo {pages}
  {1140} (\bibinfo {year} {1982})}\BibitemShut {NoStop}%
\bibitem [{\citenamefont {Pisarski}\ and\ \citenamefont
  {Wilczek}(1984)}]{Pisarski:1983ms}%
  \BibitemOpen
  \bibfield  {author} {\bibinfo {author} {\bibfnamefont {R.~D.}\ \bibnamefont
  {Pisarski}}\ and\ \bibinfo {author} {\bibfnamefont {F.}~\bibnamefont
  {Wilczek}},\ }\href {\doibase 10.1103/PhysRevD.29.338} {\bibfield  {journal}
  {\bibinfo  {journal} {Phys. Rev. D}\ }\textbf {\bibinfo {volume} {29}},\
  \bibinfo {pages} {338} (\bibinfo {year} {1984})}\BibitemShut {NoStop}%
\bibitem [{\citenamefont {Ding}\ \emph {et~al.}(2019)\citenamefont {Ding} \emph
  {et~al.}}]{Ding:2019prx}%
  \BibitemOpen
  \bibfield  {author} {\bibinfo {author} {\bibfnamefont {H.}~\bibnamefont
  {Ding}} \emph {et~al.},\ }\href {\doibase 10.1103/PhysRevLett.123.062002}
  {\bibfield  {journal} {\bibinfo  {journal} {Phys. Rev. Lett.}\ }\textbf
  {\bibinfo {volume} {123}},\ \bibinfo {pages} {062002} (\bibinfo {year}
  {2019})},\ \Eprint {http://arxiv.org/abs/1903.04801} {arXiv:1903.04801
  [hep-lat]} \BibitemShut {NoStop}%
\bibitem [{\citenamefont {Braun}\ \emph {et~al.}(2020)\citenamefont {Braun},
  \citenamefont {jie Fu}, \citenamefont {Pawlowski}, \citenamefont {Rennecke},
  \citenamefont {Rosenblüh},\ and\ \citenamefont {Yin}}]{Braun:2020ada}%
  \BibitemOpen
  \bibfield  {author} {\bibinfo {author} {\bibfnamefont {J.}~\bibnamefont
  {Braun}}, \bibinfo {author} {\bibfnamefont {W.}~\bibnamefont {jie Fu}},
  \bibinfo {author} {\bibfnamefont {J.~M.}\ \bibnamefont {Pawlowski}}, \bibinfo
  {author} {\bibfnamefont {F.}~\bibnamefont {Rennecke}}, \bibinfo {author}
  {\bibfnamefont {D.}~\bibnamefont {Rosenblüh}}, \ and\ \bibinfo {author}
  {\bibfnamefont {S.}~\bibnamefont {Yin}},\ }\href@noop {} {\enquote {\bibinfo
  {title} {Chiral susceptibility in (2+1)-flavour qcd},}\ } (\bibinfo {year}
  {2020}),\ \Eprint {http://arxiv.org/abs/2003.13112} {arXiv:2003.13112
  [hep-ph]} \BibitemShut {NoStop}%
\bibitem [{\citenamefont {Braguta}\ \emph {et~al.}(2019)\citenamefont
  {Braguta}, \citenamefont {Chernodub}, \citenamefont {Kotov}, \citenamefont
  {Molochkov},\ and\ \citenamefont {Nikolaev}}]{Braguta:2019yci}%
  \BibitemOpen
  \bibfield  {author} {\bibinfo {author} {\bibfnamefont {V.}~\bibnamefont
  {Braguta}}, \bibinfo {author} {\bibfnamefont {M.}~\bibnamefont {Chernodub}},
  \bibinfo {author} {\bibfnamefont {A.~Y.}\ \bibnamefont {Kotov}}, \bibinfo
  {author} {\bibfnamefont {A.}~\bibnamefont {Molochkov}}, \ and\ \bibinfo
  {author} {\bibfnamefont {A.}~\bibnamefont {Nikolaev}},\ }\href {\doibase
  10.1103/PhysRevD.100.114503} {\bibfield  {journal} {\bibinfo  {journal}
  {Phys. Rev. D}\ }\textbf {\bibinfo {volume} {100}},\ \bibinfo {pages}
  {114503} (\bibinfo {year} {2019})},\ \Eprint
  {http://arxiv.org/abs/1909.09547} {arXiv:1909.09547 [hep-lat]} \BibitemShut
  {NoStop}%
\bibitem [{\citenamefont {Aarts}\ \emph
  {et~al.}(2011{\natexlab{a}})\citenamefont {Aarts}, \citenamefont {Kim},
  \citenamefont {Lombardo}, \citenamefont {Oktay}, \citenamefont {Ryan},
  \citenamefont {Sinclair},\ and\ \citenamefont {Skullerud}}]{Aarts:2010ek}%
  \BibitemOpen
  \bibfield  {author} {\bibinfo {author} {\bibfnamefont {G.}~\bibnamefont
  {Aarts}}, \bibinfo {author} {\bibfnamefont {S.}~\bibnamefont {Kim}}, \bibinfo
  {author} {\bibfnamefont {M.~P.}\ \bibnamefont {Lombardo}}, \bibinfo {author}
  {\bibfnamefont {M.~B.}\ \bibnamefont {Oktay}}, \bibinfo {author}
  {\bibfnamefont {S.~M.}\ \bibnamefont {Ryan}}, \bibinfo {author}
  {\bibfnamefont {D.~K.}\ \bibnamefont {Sinclair}}, \ and\ \bibinfo {author}
  {\bibfnamefont {J.~I.}\ \bibnamefont {Skullerud}},\ }\href {\doibase
  10.1103/PhysRevLett.106.061602} {\bibfield  {journal} {\bibinfo  {journal}
  {Phys. Rev. Lett.}\ }\textbf {\bibinfo {volume} {106}},\ \bibinfo {pages}
  {061602} (\bibinfo {year} {2011}{\natexlab{a}})},\ \Eprint
  {http://arxiv.org/abs/1010.3725} {arXiv:1010.3725 [hep-lat]} \BibitemShut
  {NoStop}%
\bibitem [{\citenamefont {Aarts}\ \emph
  {et~al.}(2011{\natexlab{b}})\citenamefont {Aarts}, \citenamefont {Allton},
  \citenamefont {Kim}, \citenamefont {Lombardo}, \citenamefont {Oktay},
  \citenamefont {Ryan}, \citenamefont {Sinclair},\ and\ \citenamefont
  {Skullerud}}]{Aarts:2011sm}%
  \BibitemOpen
  \bibfield  {author} {\bibinfo {author} {\bibfnamefont {G.}~\bibnamefont
  {Aarts}}, \bibinfo {author} {\bibfnamefont {C.}~\bibnamefont {Allton}},
  \bibinfo {author} {\bibfnamefont {S.}~\bibnamefont {Kim}}, \bibinfo {author}
  {\bibfnamefont {M.~P.}\ \bibnamefont {Lombardo}}, \bibinfo {author}
  {\bibfnamefont {M.~B.}\ \bibnamefont {Oktay}}, \bibinfo {author}
  {\bibfnamefont {S.~M.}\ \bibnamefont {Ryan}}, \bibinfo {author}
  {\bibfnamefont {D.~K.}\ \bibnamefont {Sinclair}}, \ and\ \bibinfo {author}
  {\bibfnamefont {J.~I.}\ \bibnamefont {Skullerud}},\ }\href {\doibase
  10.1007/JHEP11(2011)103} {\bibfield  {journal} {\bibinfo  {journal} {JHEP}\
  }\textbf {\bibinfo {volume} {11}},\ \bibinfo {pages} {103} (\bibinfo {year}
  {2011}{\natexlab{b}})},\ \Eprint {http://arxiv.org/abs/1109.4496}
  {arXiv:1109.4496 [hep-lat]} \BibitemShut {NoStop}%
\bibitem [{\citenamefont {Aarts}\ \emph {et~al.}(2013)\citenamefont {Aarts},
  \citenamefont {Allton}, \citenamefont {Kim}, \citenamefont {Lombardo},
  \citenamefont {Ryan},\ and\ \citenamefont {Skullerud}}]{Aarts:2013kaa}%
  \BibitemOpen
  \bibfield  {author} {\bibinfo {author} {\bibfnamefont {G.}~\bibnamefont
  {Aarts}}, \bibinfo {author} {\bibfnamefont {C.}~\bibnamefont {Allton}},
  \bibinfo {author} {\bibfnamefont {S.}~\bibnamefont {Kim}}, \bibinfo {author}
  {\bibfnamefont {M.~P.}\ \bibnamefont {Lombardo}}, \bibinfo {author}
  {\bibfnamefont {S.~M.}\ \bibnamefont {Ryan}}, \ and\ \bibinfo {author}
  {\bibfnamefont {J.~I.}\ \bibnamefont {Skullerud}},\ }\href {\doibase
  10.1007/JHEP12(2013)064} {\bibfield  {journal} {\bibinfo  {journal} {JHEP}\
  }\textbf {\bibinfo {volume} {12}},\ \bibinfo {pages} {064} (\bibinfo {year}
  {2013})},\ \Eprint {http://arxiv.org/abs/1310.5467} {arXiv:1310.5467
  [hep-lat]} \BibitemShut {NoStop}%
\bibitem [{\citenamefont {Aarts}\ \emph {et~al.}(2014)\citenamefont {Aarts},
  \citenamefont {Allton}, \citenamefont {Harris}, \citenamefont {Kim},
  \citenamefont {Lombardo}, \citenamefont {Ryan},\ and\ \citenamefont
  {Skullerud}}]{Aarts:2014cda}%
  \BibitemOpen
  \bibfield  {author} {\bibinfo {author} {\bibfnamefont {G.}~\bibnamefont
  {Aarts}}, \bibinfo {author} {\bibfnamefont {C.}~\bibnamefont {Allton}},
  \bibinfo {author} {\bibfnamefont {T.}~\bibnamefont {Harris}}, \bibinfo
  {author} {\bibfnamefont {S.}~\bibnamefont {Kim}}, \bibinfo {author}
  {\bibfnamefont {M.~P.}\ \bibnamefont {Lombardo}}, \bibinfo {author}
  {\bibfnamefont {S.~M.}\ \bibnamefont {Ryan}}, \ and\ \bibinfo {author}
  {\bibfnamefont {J.-I.}\ \bibnamefont {Skullerud}},\ }\href {\doibase
  10.1007/JHEP07(2014)097} {\bibfield  {journal} {\bibinfo  {journal} {JHEP}\
  }\textbf {\bibinfo {volume} {07}},\ \bibinfo {pages} {097} (\bibinfo {year}
  {2014})},\ \Eprint {http://arxiv.org/abs/1402.6210} {arXiv:1402.6210
  [hep-lat]} \BibitemShut {NoStop}%
\bibitem [{\citenamefont {Kelly}\ \emph {et~al.}(2018)\citenamefont {Kelly},
  \citenamefont {Rothkopf},\ and\ \citenamefont {Skullerud}}]{Kelly:2018hsi}%
  \BibitemOpen
  \bibfield  {author} {\bibinfo {author} {\bibfnamefont {A.}~\bibnamefont
  {Kelly}}, \bibinfo {author} {\bibfnamefont {A.}~\bibnamefont {Rothkopf}}, \
  and\ \bibinfo {author} {\bibfnamefont {J.-I.}\ \bibnamefont {Skullerud}},\
  }\href {\doibase 10.1103/PhysRevD.97.114509} {\bibfield  {journal} {\bibinfo
  {journal} {Phys. Rev.}\ }\textbf {\bibinfo {volume} {D97}},\ \bibinfo {pages}
  {114509} (\bibinfo {year} {2018})},\ \Eprint
  {http://arxiv.org/abs/1802.00667} {arXiv:1802.00667 [hep-lat]} \BibitemShut
  {NoStop}%
\bibitem [{\citenamefont {Aarts}\ \emph
  {et~al.}(2015{\natexlab{a}})\citenamefont {Aarts}, \citenamefont {Allton},
  \citenamefont {Hands}, \citenamefont {J{\"a}ger}, \citenamefont {Praki},\
  and\ \citenamefont {Skullerud}}]{Aarts:2015mma}%
  \BibitemOpen
  \bibfield  {author} {\bibinfo {author} {\bibfnamefont {G.}~\bibnamefont
  {Aarts}}, \bibinfo {author} {\bibfnamefont {C.}~\bibnamefont {Allton}},
  \bibinfo {author} {\bibfnamefont {S.}~\bibnamefont {Hands}}, \bibinfo
  {author} {\bibfnamefont {B.}~\bibnamefont {J{\"a}ger}}, \bibinfo {author}
  {\bibfnamefont {C.}~\bibnamefont {Praki}}, \ and\ \bibinfo {author}
  {\bibfnamefont {J.-I.}\ \bibnamefont {Skullerud}},\ }\href {\doibase
  10.1103/PhysRevD.92.014503} {\bibfield  {journal} {\bibinfo  {journal} {Phys.
  Rev.}\ }\textbf {\bibinfo {volume} {D92}},\ \bibinfo {pages} {014503}
  (\bibinfo {year} {2015}{\natexlab{a}})},\ \Eprint
  {http://arxiv.org/abs/1502.03603} {arXiv:1502.03603 [hep-lat]} \BibitemShut
  {NoStop}%
\bibitem [{\citenamefont {Aarts}\ \emph {et~al.}(2017)\citenamefont {Aarts},
  \citenamefont {Allton}, \citenamefont {De~Boni}, \citenamefont {Hands},
  \citenamefont {J{\"a}ger}, \citenamefont {Praki},\ and\ \citenamefont
  {Skullerud}}]{Aarts:2017rrl}%
  \BibitemOpen
  \bibfield  {author} {\bibinfo {author} {\bibfnamefont {G.}~\bibnamefont
  {Aarts}}, \bibinfo {author} {\bibfnamefont {C.}~\bibnamefont {Allton}},
  \bibinfo {author} {\bibfnamefont {D.}~\bibnamefont {De~Boni}}, \bibinfo
  {author} {\bibfnamefont {S.}~\bibnamefont {Hands}}, \bibinfo {author}
  {\bibfnamefont {B.}~\bibnamefont {J{\"a}ger}}, \bibinfo {author}
  {\bibfnamefont {C.}~\bibnamefont {Praki}}, \ and\ \bibinfo {author}
  {\bibfnamefont {J.-I.}\ \bibnamefont {Skullerud}},\ }\href {\doibase
  10.1007/JHEP06(2017)034} {\bibfield  {journal} {\bibinfo  {journal} {JHEP}\
  }\textbf {\bibinfo {volume} {06}},\ \bibinfo {pages} {034} (\bibinfo {year}
  {2017})},\ \Eprint {http://arxiv.org/abs/1703.09246} {arXiv:1703.09246
  [hep-lat]} \BibitemShut {NoStop}%
\bibitem [{\citenamefont {Aarts}\ \emph
  {et~al.}(2019{\natexlab{a}})\citenamefont {Aarts}, \citenamefont {Allton},
  \citenamefont {De~Boni},\ and\ \citenamefont {J{\"a}ger}}]{Aarts:2018glk}%
  \BibitemOpen
  \bibfield  {author} {\bibinfo {author} {\bibfnamefont {G.}~\bibnamefont
  {Aarts}}, \bibinfo {author} {\bibfnamefont {C.}~\bibnamefont {Allton}},
  \bibinfo {author} {\bibfnamefont {D.}~\bibnamefont {De~Boni}}, \ and\
  \bibinfo {author} {\bibfnamefont {B.}~\bibnamefont {J{\"a}ger}},\ }\href
  {\doibase 10.1103/PhysRevD.99.074503} {\bibfield  {journal} {\bibinfo
  {journal} {Phys. Rev.}\ }\textbf {\bibinfo {volume} {D99}},\ \bibinfo {pages}
  {074503} (\bibinfo {year} {2019}{\natexlab{a}})},\ \Eprint
  {http://arxiv.org/abs/1812.07393} {arXiv:1812.07393 [hep-lat]} \BibitemShut
  {NoStop}%
\bibitem [{\citenamefont {Bors\'anyi}\ \emph {et~al.}(2012)\citenamefont
  {Bors\'anyi}, \citenamefont {Durr}, \citenamefont {Fodor}, \citenamefont
  {Hoelbling}, \citenamefont {Katz}, \citenamefont {Krieg}, \citenamefont
  {Nogradi}, \citenamefont {Szabo}, \citenamefont {Toth},\ and\ \citenamefont
  {Trombitas}}]{Borsanyi:2012uq}%
  \BibitemOpen
  \bibfield  {author} {\bibinfo {author} {\bibfnamefont {S.}~\bibnamefont
  {Bors\'anyi}}, \bibinfo {author} {\bibfnamefont {S.}~\bibnamefont {Durr}},
  \bibinfo {author} {\bibfnamefont {Z.}~\bibnamefont {Fodor}}, \bibinfo
  {author} {\bibfnamefont {C.}~\bibnamefont {Hoelbling}}, \bibinfo {author}
  {\bibfnamefont {S.~D.}\ \bibnamefont {Katz}}, \bibinfo {author}
  {\bibfnamefont {S.}~\bibnamefont {Krieg}}, \bibinfo {author} {\bibfnamefont
  {D.}~\bibnamefont {Nogradi}}, \bibinfo {author} {\bibfnamefont {K.~K.}\
  \bibnamefont {Szabo}}, \bibinfo {author} {\bibfnamefont {B.~C.}\ \bibnamefont
  {Toth}}, \ and\ \bibinfo {author} {\bibfnamefont {N.}~\bibnamefont
  {Trombitas}},\ }\href {\doibase 10.1007/JHEP08(2012)126} {\bibfield
  {journal} {\bibinfo  {journal} {JHEP}\ }\textbf {\bibinfo {volume} {08}},\
  \bibinfo {pages} {126} (\bibinfo {year} {2012})},\ \Eprint
  {http://arxiv.org/abs/1205.0440} {arXiv:1205.0440 [hep-lat]} \BibitemShut
  {NoStop}%
\bibitem [{\citenamefont {Bors{\'a}nyi}\ \emph {et~al.}(2015)\citenamefont
  {Bors{\'a}nyi}, \citenamefont {Durr}, \citenamefont {Fodor}, \citenamefont
  {Holbling}, \citenamefont {Katz}, \citenamefont {Krieg}, \citenamefont
  {Nogradi}, \citenamefont {Szabo}, \citenamefont {Toth},\ and\ \citenamefont
  {Trombitas}}]{Borsanyi:2015waa}%
  \BibitemOpen
  \bibfield  {author} {\bibinfo {author} {\bibfnamefont {S.}~\bibnamefont
  {Bors{\'a}nyi}}, \bibinfo {author} {\bibfnamefont {S.}~\bibnamefont {Durr}},
  \bibinfo {author} {\bibfnamefont {Z.}~\bibnamefont {Fodor}}, \bibinfo
  {author} {\bibfnamefont {C.}~\bibnamefont {Holbling}}, \bibinfo {author}
  {\bibfnamefont {S.~D.}\ \bibnamefont {Katz}}, \bibinfo {author}
  {\bibfnamefont {S.}~\bibnamefont {Krieg}}, \bibinfo {author} {\bibfnamefont
  {D.}~\bibnamefont {Nogradi}}, \bibinfo {author} {\bibfnamefont {K.~K.}\
  \bibnamefont {Szabo}}, \bibinfo {author} {\bibfnamefont {B.~C.}\ \bibnamefont
  {Toth}}, \ and\ \bibinfo {author} {\bibfnamefont {N.}~\bibnamefont
  {Trombitas}},\ }\href {\doibase 10.1103/PhysRevD.92.014505} {\bibfield
  {journal} {\bibinfo  {journal} {Phys. Rev.}\ }\textbf {\bibinfo {volume}
  {D92}},\ \bibinfo {pages} {014505} (\bibinfo {year} {2015})},\ \Eprint
  {http://arxiv.org/abs/1504.03676} {arXiv:1504.03676 [hep-lat]} \BibitemShut
  {NoStop}%
\bibitem [{\citenamefont {Umeda}\ \emph {et~al.}(2012)\citenamefont {Umeda},
  \citenamefont {Aoki}, \citenamefont {Ejiri}, \citenamefont {Hatsuda},
  \citenamefont {Kanaya}, \citenamefont {Maezawa},\ and\ \citenamefont
  {Ohno}}]{Umeda:2012er}%
  \BibitemOpen
  \bibfield  {author} {\bibinfo {author} {\bibfnamefont {T.}~\bibnamefont
  {Umeda}}, \bibinfo {author} {\bibfnamefont {S.}~\bibnamefont {Aoki}},
  \bibinfo {author} {\bibfnamefont {S.}~\bibnamefont {Ejiri}}, \bibinfo
  {author} {\bibfnamefont {T.}~\bibnamefont {Hatsuda}}, \bibinfo {author}
  {\bibfnamefont {K.}~\bibnamefont {Kanaya}}, \bibinfo {author} {\bibfnamefont
  {Y.}~\bibnamefont {Maezawa}}, \ and\ \bibinfo {author} {\bibfnamefont
  {H.}~\bibnamefont {Ohno}} (\bibinfo {collaboration} {WHOT-QCD}),\ }\href
  {\doibase 10.1103/PhysRevD.85.094508} {\bibfield  {journal} {\bibinfo
  {journal} {Phys. Rev. D}\ }\textbf {\bibinfo {volume} {85}},\ \bibinfo
  {pages} {094508} (\bibinfo {year} {2012})},\ \Eprint
  {http://arxiv.org/abs/1202.4719} {arXiv:1202.4719 [hep-lat]} \BibitemShut
  {NoStop}%
\bibitem [{\citenamefont {Taniguchi}\ \emph {et~al.}(2017)\citenamefont
  {Taniguchi}, \citenamefont {Ejiri}, \citenamefont {Iwami}, \citenamefont
  {Kanaya}, \citenamefont {Kitazawa}, \citenamefont {Suzuki}, \citenamefont
  {Umeda},\ and\ \citenamefont {Wakabayashi}}]{Taniguchi:2016ofw}%
  \BibitemOpen
  \bibfield  {author} {\bibinfo {author} {\bibfnamefont {Y.}~\bibnamefont
  {Taniguchi}}, \bibinfo {author} {\bibfnamefont {S.}~\bibnamefont {Ejiri}},
  \bibinfo {author} {\bibfnamefont {R.}~\bibnamefont {Iwami}}, \bibinfo
  {author} {\bibfnamefont {K.}~\bibnamefont {Kanaya}}, \bibinfo {author}
  {\bibfnamefont {M.}~\bibnamefont {Kitazawa}}, \bibinfo {author}
  {\bibfnamefont {H.}~\bibnamefont {Suzuki}}, \bibinfo {author} {\bibfnamefont
  {T.}~\bibnamefont {Umeda}}, \ and\ \bibinfo {author} {\bibfnamefont
  {N.}~\bibnamefont {Wakabayashi}},\ }\href {\doibase
  10.1103/PhysRevD.96.014509} {\bibfield  {journal} {\bibinfo  {journal} {Phys.
  Rev. D}\ }\textbf {\bibinfo {volume} {96}},\ \bibinfo {pages} {014509}
  (\bibinfo {year} {2017})},\ \bibinfo {note} {[Erratum: Phys.Rev.D 99, 059904
  (2019)]},\ \Eprint {http://arxiv.org/abs/1609.01417} {arXiv:1609.01417
  [hep-lat]} \BibitemShut {NoStop}%
\bibitem [{\citenamefont {Taniguchi}\ \emph {et~al.}(2020)\citenamefont
  {Taniguchi}, \citenamefont {Ejiri}, \citenamefont {Kanaya}, \citenamefont
  {Kitazawa}, \citenamefont {Suzuki},\ and\ \citenamefont
  {Umeda}}]{Taniguchi:2020mgg}%
  \BibitemOpen
  \bibfield  {author} {\bibinfo {author} {\bibfnamefont {Y.}~\bibnamefont
  {Taniguchi}}, \bibinfo {author} {\bibfnamefont {S.}~\bibnamefont {Ejiri}},
  \bibinfo {author} {\bibfnamefont {K.}~\bibnamefont {Kanaya}}, \bibinfo
  {author} {\bibfnamefont {M.}~\bibnamefont {Kitazawa}}, \bibinfo {author}
  {\bibfnamefont {H.}~\bibnamefont {Suzuki}}, \ and\ \bibinfo {author}
  {\bibfnamefont {T.}~\bibnamefont {Umeda}},\ }\href@noop {} {\enquote
  {\bibinfo {title} {Nf=2+1 qcd thermodynamics with gradient flow using
  two-loop matching coefficients},}\ } (\bibinfo {year} {2020}),\ \Eprint
  {http://arxiv.org/abs/2005.00251} {arXiv:2005.00251 [hep-lat]} \BibitemShut
  {NoStop}%
\bibitem [{\citenamefont {Kanaya}\ \emph {et~al.}(2019)\citenamefont {Kanaya},
  \citenamefont {Baba}, \citenamefont {Suzuki}, \citenamefont {Ejiri},
  \citenamefont {Kitazawa}, \citenamefont {Suzuki}, \citenamefont {Taniguchi},\
  and\ \citenamefont {Umeda}}]{Kanaya:2019okb}%
  \BibitemOpen
  \bibfield  {author} {\bibinfo {author} {\bibfnamefont {K.}~\bibnamefont
  {Kanaya}}, \bibinfo {author} {\bibfnamefont {A.}~\bibnamefont {Baba}},
  \bibinfo {author} {\bibfnamefont {A.}~\bibnamefont {Suzuki}}, \bibinfo
  {author} {\bibfnamefont {S.}~\bibnamefont {Ejiri}}, \bibinfo {author}
  {\bibfnamefont {M.}~\bibnamefont {Kitazawa}}, \bibinfo {author}
  {\bibfnamefont {H.}~\bibnamefont {Suzuki}}, \bibinfo {author} {\bibfnamefont
  {Y.}~\bibnamefont {Taniguchi}}, \ and\ \bibinfo {author} {\bibfnamefont
  {T.}~\bibnamefont {Umeda}},\ }\href@noop {} {\bibfield  {journal} {\bibinfo
  {journal} {PoS}\ }\textbf {\bibinfo {volume} {LATTICE2019}},\ \bibinfo
  {pages} {088} (\bibinfo {year} {2019})},\ \Eprint
  {http://arxiv.org/abs/1910.13036} {arXiv:1910.13036 [hep-lat]} \BibitemShut
  {NoStop}%
\bibitem [{\citenamefont {Burger}\ \emph {et~al.}(2018)\citenamefont {Burger},
  \citenamefont {Ilgenfritz}, \citenamefont {Lombardo},\ and\ \citenamefont
  {Trunin}}]{Burger:2018fvb}%
  \BibitemOpen
  \bibfield  {author} {\bibinfo {author} {\bibfnamefont {F.}~\bibnamefont
  {Burger}}, \bibinfo {author} {\bibfnamefont {E.-M.}\ \bibnamefont
  {Ilgenfritz}}, \bibinfo {author} {\bibfnamefont {M.~P.}\ \bibnamefont
  {Lombardo}}, \ and\ \bibinfo {author} {\bibfnamefont {A.}~\bibnamefont
  {Trunin}},\ }\href {\doibase 10.1103/PhysRevD.98.094501} {\bibfield
  {journal} {\bibinfo  {journal} {Phys. Rev.}\ }\textbf {\bibinfo {volume}
  {D98}},\ \bibinfo {pages} {094501} (\bibinfo {year} {2018})},\ \Eprint
  {http://arxiv.org/abs/1805.06001} {arXiv:1805.06001 [hep-lat]} \BibitemShut
  {NoStop}%
\bibitem [{\citenamefont {Kotov}\ \emph {et~al.}(2020)\citenamefont {Kotov},
  \citenamefont {Lombardo},\ and\ \citenamefont {Trunin}}]{Kotov:2020hzm}%
  \BibitemOpen
  \bibfield  {author} {\bibinfo {author} {\bibfnamefont {A.~Y.}\ \bibnamefont
  {Kotov}}, \bibinfo {author} {\bibfnamefont {M.~P.}\ \bibnamefont {Lombardo}},
  \ and\ \bibinfo {author} {\bibfnamefont {A.~M.}\ \bibnamefont {Trunin}},\
  }\href@noop {} {\enquote {\bibinfo {title} {Finite temperature qcd with
  $n_f=2+1+1$ wilson twisted mass fermions at physical pion, strange and charm
  masses},}\ } (\bibinfo {year} {2020}),\ \Eprint
  {http://arxiv.org/abs/2004.07122} {arXiv:2004.07122 [hep-lat]} \BibitemShut
  {NoStop}%
\bibitem [{\citenamefont {Aarts}\ \emph {et~al.}(2018)\citenamefont {Aarts},
  \citenamefont {Allton}, \citenamefont {Glesaaen}, \citenamefont {Hands},
  \citenamefont {J{\"a}ger},\ and\ \citenamefont {Skullerud}}]{Aarts:2018haw}%
  \BibitemOpen
  \bibfield  {author} {\bibinfo {author} {\bibfnamefont {G.}~\bibnamefont
  {Aarts}}, \bibinfo {author} {\bibfnamefont {C.}~\bibnamefont {Allton}},
  \bibinfo {author} {\bibfnamefont {J.}~\bibnamefont {Glesaaen}}, \bibinfo
  {author} {\bibfnamefont {S.}~\bibnamefont {Hands}}, \bibinfo {author}
  {\bibfnamefont {B.}~\bibnamefont {J{\"a}ger}}, \ and\ \bibinfo {author}
  {\bibfnamefont {J.}~\bibnamefont {Skullerud}},\ }\href {\doibase
  10.22323/1.334.0183} {\bibfield  {journal} {\bibinfo  {journal} {PoS}\
  }\textbf {\bibinfo {volume} {LATTICE2018}},\ \bibinfo {pages} {183} (\bibinfo
  {year} {2018})},\ \Eprint {http://arxiv.org/abs/1812.08151} {arXiv:1812.08151
  [hep-lat]} \BibitemShut {NoStop}%
\bibitem [{\citenamefont {Aarts}\ \emph
  {et~al.}(2019{\natexlab{b}})\citenamefont {Aarts} \emph
  {et~al.}}]{Aarts:2019hrg}%
  \BibitemOpen
  \bibfield  {author} {\bibinfo {author} {\bibfnamefont {G.}~\bibnamefont
  {Aarts}} \emph {et~al.},\ }\href@noop {} {\bibfield  {journal} {\bibinfo
  {journal} {PoS}\ }\textbf {\bibinfo {volume} {LATTICE2019}},\ \bibinfo
  {pages} {075} (\bibinfo {year} {2019}{\natexlab{b}})},\ \Eprint
  {http://arxiv.org/abs/1912.09827} {arXiv:1912.09827 [hep-lat]} \BibitemShut
  {NoStop}%
\bibitem [{\citenamefont {Wilson}\ \emph {et~al.}(2019)\citenamefont {Wilson},
  \citenamefont {Briceno}, \citenamefont {Dudek}, \citenamefont {Edwards},\
  and\ \citenamefont {Thomas}}]{Wilson:2019wfr}%
  \BibitemOpen
  \bibfield  {author} {\bibinfo {author} {\bibfnamefont {D.~J.}\ \bibnamefont
  {Wilson}}, \bibinfo {author} {\bibfnamefont {R.~A.}\ \bibnamefont {Briceno}},
  \bibinfo {author} {\bibfnamefont {J.~J.}\ \bibnamefont {Dudek}}, \bibinfo
  {author} {\bibfnamefont {R.~G.}\ \bibnamefont {Edwards}}, \ and\ \bibinfo
  {author} {\bibfnamefont {C.~E.}\ \bibnamefont {Thomas}},\ }\href {\doibase
  10.1103/PhysRevLett.123.042002} {\bibfield  {journal} {\bibinfo  {journal}
  {Phys. Rev. Lett.}\ }\textbf {\bibinfo {volume} {123}},\ \bibinfo {pages}
  {042002} (\bibinfo {year} {2019})},\ \Eprint
  {http://arxiv.org/abs/1904.03188} {arXiv:1904.03188 [hep-lat]} \BibitemShut
  {NoStop}%
\bibitem [{\citenamefont {Cheung}\ \emph {et~al.}(2016)\citenamefont {Cheung},
  \citenamefont {O'Hara}, \citenamefont {Moir}, \citenamefont {Peardon},
  \citenamefont {Ryan}, \citenamefont {Thomas},\ and\ \citenamefont
  {Tims}}]{Cheung:2016bym}%
  \BibitemOpen
  \bibfield  {author} {\bibinfo {author} {\bibfnamefont {G.~K.~C.}\
  \bibnamefont {Cheung}}, \bibinfo {author} {\bibfnamefont {C.}~\bibnamefont
  {O'Hara}}, \bibinfo {author} {\bibfnamefont {G.}~\bibnamefont {Moir}},
  \bibinfo {author} {\bibfnamefont {M.}~\bibnamefont {Peardon}}, \bibinfo
  {author} {\bibfnamefont {S.~M.}\ \bibnamefont {Ryan}}, \bibinfo {author}
  {\bibfnamefont {C.~E.}\ \bibnamefont {Thomas}}, \ and\ \bibinfo {author}
  {\bibfnamefont {D.}~\bibnamefont {Tims}} (\bibinfo {collaboration} {Hadron
  Spectrum}),\ }\href {\doibase 10.1007/JHEP12(2016)089} {\bibfield  {journal}
  {\bibinfo  {journal} {JHEP}\ }\textbf {\bibinfo {volume} {12}},\ \bibinfo
  {pages} {089} (\bibinfo {year} {2016})},\ \Eprint
  {http://arxiv.org/abs/1610.01073} {arXiv:1610.01073 [hep-lat]} \BibitemShut
  {NoStop}%
\bibitem [{\citenamefont {Edwards}\ \emph {et~al.}(2008)\citenamefont
  {Edwards}, \citenamefont {Joo},\ and\ \citenamefont {Lin}}]{Edwards:2008ja}%
  \BibitemOpen
  \bibfield  {author} {\bibinfo {author} {\bibfnamefont {R.~G.}\ \bibnamefont
  {Edwards}}, \bibinfo {author} {\bibfnamefont {B.}~\bibnamefont {Joo}}, \ and\
  \bibinfo {author} {\bibfnamefont {H.-W.}\ \bibnamefont {Lin}},\ }\href
  {\doibase 10.1103/PhysRevD.78.054501} {\bibfield  {journal} {\bibinfo
  {journal} {Phys. Rev.}\ }\textbf {\bibinfo {volume} {D78}},\ \bibinfo {pages}
  {054501} (\bibinfo {year} {2008})},\ \Eprint {http://arxiv.org/abs/0803.3960}
  {arXiv:0803.3960 [hep-lat]} \BibitemShut {NoStop}%
\bibitem [{\citenamefont {Lin}\ \emph {et~al.}(2009)\citenamefont {Lin} \emph
  {et~al.}}]{Lin:2008pr}%
  \BibitemOpen
  \bibfield  {author} {\bibinfo {author} {\bibfnamefont {H.-W.}\ \bibnamefont
  {Lin}} \emph {et~al.} (\bibinfo {collaboration} {Hadron Spectrum}),\ }\href
  {\doibase 10.1103/PhysRevD.79.034502} {\bibfield  {journal} {\bibinfo
  {journal} {Phys. Rev.}\ }\textbf {\bibinfo {volume} {D79}},\ \bibinfo {pages}
  {034502} (\bibinfo {year} {2009})},\ \Eprint {http://arxiv.org/abs/0810.3588}
  {arXiv:0810.3588 [hep-lat]} \BibitemShut {NoStop}%
\bibitem [{\citenamefont {Aarts}\ \emph
  {et~al.}(2015{\natexlab{b}})\citenamefont {Aarts}, \citenamefont {Allton},
  \citenamefont {Amato}, \citenamefont {Giudice}, \citenamefont {Hands},\ and\
  \citenamefont {Skullerud}}]{Aarts:2014nba}%
  \BibitemOpen
  \bibfield  {author} {\bibinfo {author} {\bibfnamefont {G.}~\bibnamefont
  {Aarts}}, \bibinfo {author} {\bibfnamefont {C.}~\bibnamefont {Allton}},
  \bibinfo {author} {\bibfnamefont {A.}~\bibnamefont {Amato}}, \bibinfo
  {author} {\bibfnamefont {P.}~\bibnamefont {Giudice}}, \bibinfo {author}
  {\bibfnamefont {S.}~\bibnamefont {Hands}}, \ and\ \bibinfo {author}
  {\bibfnamefont {J.-I.}\ \bibnamefont {Skullerud}},\ }\href {\doibase
  10.1007/JHEP02(2015)186} {\bibfield  {journal} {\bibinfo  {journal} {JHEP}\
  }\textbf {\bibinfo {volume} {02}},\ \bibinfo {pages} {186} (\bibinfo {year}
  {2015}{\natexlab{b}})},\ \Eprint {http://arxiv.org/abs/1412.6411}
  {arXiv:1412.6411 [hep-lat]} \BibitemShut {NoStop}%
\bibitem [{\citenamefont {Dudek}\ \emph {et~al.}(2012)\citenamefont {Dudek},
  \citenamefont {Edwards},\ and\ \citenamefont {Thomas}}]{Dudek:2012gj}%
  \BibitemOpen
  \bibfield  {author} {\bibinfo {author} {\bibfnamefont {J.~J.}\ \bibnamefont
  {Dudek}}, \bibinfo {author} {\bibfnamefont {R.~G.}\ \bibnamefont {Edwards}},
  \ and\ \bibinfo {author} {\bibfnamefont {C.~E.}\ \bibnamefont {Thomas}},\
  }\href {\doibase 10.1103/PhysRevD.86.034031} {\bibfield  {journal} {\bibinfo
  {journal} {Phys. Rev. D}\ }\textbf {\bibinfo {volume} {86}},\ \bibinfo
  {pages} {034031} (\bibinfo {year} {2012})},\ \Eprint
  {http://arxiv.org/abs/1203.6041} {arXiv:1203.6041 [hep-ph]} \BibitemShut
  {NoStop}%
\bibitem [{\citenamefont {Wilson}\ \emph {et~al.}(2015)\citenamefont {Wilson},
  \citenamefont {Briceno}, \citenamefont {Dudek}, \citenamefont {Edwards},\
  and\ \citenamefont {Thomas}}]{Wilson:2015dqa}%
  \BibitemOpen
  \bibfield  {author} {\bibinfo {author} {\bibfnamefont {D.~J.}\ \bibnamefont
  {Wilson}}, \bibinfo {author} {\bibfnamefont {R.~A.}\ \bibnamefont {Briceno}},
  \bibinfo {author} {\bibfnamefont {J.~J.}\ \bibnamefont {Dudek}}, \bibinfo
  {author} {\bibfnamefont {R.~G.}\ \bibnamefont {Edwards}}, \ and\ \bibinfo
  {author} {\bibfnamefont {C.~E.}\ \bibnamefont {Thomas}},\ }\href {\doibase
  10.1103/PhysRevD.92.094502} {\bibfield  {journal} {\bibinfo  {journal} {Phys.
  Rev.}\ }\textbf {\bibinfo {volume} {D92}},\ \bibinfo {pages} {094502}
  (\bibinfo {year} {2015})},\ \Eprint {http://arxiv.org/abs/1507.02599}
  {arXiv:1507.02599 [hep-ph]} \BibitemShut {NoStop}%
\bibitem [{\citenamefont {Bazavov}\ \emph {et~al.}(2016)\citenamefont
  {Bazavov}, \citenamefont {Brambilla}, \citenamefont {Ding}, \citenamefont
  {Petreczky}, \citenamefont {Schadler}, \citenamefont {Vairo},\ and\
  \citenamefont {Weber}}]{Bazavov:2016uvm}%
  \BibitemOpen
  \bibfield  {author} {\bibinfo {author} {\bibfnamefont {A.}~\bibnamefont
  {Bazavov}}, \bibinfo {author} {\bibfnamefont {N.}~\bibnamefont {Brambilla}},
  \bibinfo {author} {\bibfnamefont {H.~T.}\ \bibnamefont {Ding}}, \bibinfo
  {author} {\bibfnamefont {P.}~\bibnamefont {Petreczky}}, \bibinfo {author}
  {\bibfnamefont {H.~P.}\ \bibnamefont {Schadler}}, \bibinfo {author}
  {\bibfnamefont {A.}~\bibnamefont {Vairo}}, \ and\ \bibinfo {author}
  {\bibfnamefont {J.}~\bibnamefont {Weber}},\ }\href {\doibase
  10.1103/PhysRevD.93.114502} {\bibfield  {journal} {\bibinfo  {journal} {Phys.
  Rev. D}\ }\textbf {\bibinfo {volume} {93}},\ \bibinfo {pages} {114502}
  (\bibinfo {year} {2016})},\ \Eprint {http://arxiv.org/abs/1603.06637}
  {arXiv:1603.06637 [hep-lat]} \BibitemShut {NoStop}%
\bibitem [{\citenamefont {Weber}(2016)}]{Weber:2016fgn}%
  \BibitemOpen
  \bibfield  {author} {\bibinfo {author} {\bibfnamefont {J.~H.}\ \bibnamefont
  {Weber}} (\bibinfo {collaboration} {TUMQCD}),\ }\href {\doibase
  10.1142/S0217732316300408} {\bibfield  {journal} {\bibinfo  {journal} {Mod.
  Phys. Lett. A}\ }\textbf {\bibinfo {volume} {31}},\ \bibinfo {pages}
  {1630040} (\bibinfo {year} {2016})},\ \Eprint
  {http://arxiv.org/abs/1606.06193} {arXiv:1606.06193 [hep-lat]} \BibitemShut
  {NoStop}%
\bibitem [{\citenamefont {Giudice}\ \emph {et~al.}(2014)\citenamefont
  {Giudice}, \citenamefont {Aarts}, \citenamefont {Allton}, \citenamefont
  {Amato}, \citenamefont {Hands},\ and\ \citenamefont
  {Skullerud}}]{Giudice:2013fza}%
  \BibitemOpen
  \bibfield  {author} {\bibinfo {author} {\bibfnamefont {P.}~\bibnamefont
  {Giudice}}, \bibinfo {author} {\bibfnamefont {G.}~\bibnamefont {Aarts}},
  \bibinfo {author} {\bibfnamefont {C.}~\bibnamefont {Allton}}, \bibinfo
  {author} {\bibfnamefont {A.}~\bibnamefont {Amato}}, \bibinfo {author}
  {\bibfnamefont {S.}~\bibnamefont {Hands}}, \ and\ \bibinfo {author}
  {\bibfnamefont {J.-I.}\ \bibnamefont {Skullerud}},\ }\href {\doibase
  10.22323/1.187.0492} {\bibfield  {journal} {\bibinfo  {journal} {PoS}\
  }\textbf {\bibinfo {volume} {LATTICE2013}},\ \bibinfo {pages} {492} (\bibinfo
  {year} {2014})},\ \Eprint {http://arxiv.org/abs/1309.6253} {arXiv:1309.6253
  [hep-lat]} \BibitemShut {NoStop}%
\bibitem [{\citenamefont {Giusti}\ \emph {et~al.}(1999)\citenamefont {Giusti},
  \citenamefont {Rapuano}, \citenamefont {Talevi},\ and\ \citenamefont
  {Vladikas}}]{Giusti:1998wy}%
  \BibitemOpen
  \bibfield  {author} {\bibinfo {author} {\bibfnamefont {L.}~\bibnamefont
  {Giusti}}, \bibinfo {author} {\bibfnamefont {F.}~\bibnamefont {Rapuano}},
  \bibinfo {author} {\bibfnamefont {M.}~\bibnamefont {Talevi}}, \ and\ \bibinfo
  {author} {\bibfnamefont {A.}~\bibnamefont {Vladikas}},\ }\href {\doibase
  10.1016/S0550-3213(98)00659-2} {\bibfield  {journal} {\bibinfo  {journal}
  {Nucl. Phys.}\ }\textbf {\bibinfo {volume} {B538}},\ \bibinfo {pages} {249}
  (\bibinfo {year} {1999})},\ \Eprint {http://arxiv.org/abs/hep-lat/9807014}
  {arXiv:hep-lat/9807014 [hep-lat]} \BibitemShut {NoStop}%
\bibitem [{\citenamefont {Datta}\ \emph {et~al.}(2013)\citenamefont {Datta},
  \citenamefont {Gupta}, \citenamefont {Padmanath}, \citenamefont {Maiti},\
  and\ \citenamefont {Mathur}}]{Datta:2012fz}%
  \BibitemOpen
  \bibfield  {author} {\bibinfo {author} {\bibfnamefont {S.}~\bibnamefont
  {Datta}}, \bibinfo {author} {\bibfnamefont {S.}~\bibnamefont {Gupta}},
  \bibinfo {author} {\bibfnamefont {M.}~\bibnamefont {Padmanath}}, \bibinfo
  {author} {\bibfnamefont {J.}~\bibnamefont {Maiti}}, \ and\ \bibinfo {author}
  {\bibfnamefont {N.}~\bibnamefont {Mathur}},\ }\href {\doibase
  10.1007/JHEP02(2013)145} {\bibfield  {journal} {\bibinfo  {journal} {JHEP}\
  }\textbf {\bibinfo {volume} {02}},\ \bibinfo {pages} {145} (\bibinfo {year}
  {2013})},\ \Eprint {http://arxiv.org/abs/1212.2927} {arXiv:1212.2927
  [hep-lat]} \BibitemShut {NoStop}%
\bibitem [{\citenamefont {Engels}\ and\ \citenamefont
  {Karsch}(2012)}]{Engels:2011km}%
  \BibitemOpen
  \bibfield  {author} {\bibinfo {author} {\bibfnamefont {J.}~\bibnamefont
  {Engels}}\ and\ \bibinfo {author} {\bibfnamefont {F.}~\bibnamefont
  {Karsch}},\ }\href {\doibase 10.1103/PhysRevD.85.094506} {\bibfield
  {journal} {\bibinfo  {journal} {Phys. Rev. D}\ }\textbf {\bibinfo {volume}
  {85}},\ \bibinfo {pages} {094506} (\bibinfo {year} {2012})},\ \Eprint
  {http://arxiv.org/abs/1105.0584} {arXiv:1105.0584 [hep-lat]} \BibitemShut
  {NoStop}%
\bibitem [{\citenamefont {Bazavov}\ \emph {et~al.}(2019)\citenamefont {Bazavov}
  \emph {et~al.}}]{Bazavov:2018mes}%
  \BibitemOpen
  \bibfield  {author} {\bibinfo {author} {\bibfnamefont {A.}~\bibnamefont
  {Bazavov}} \emph {et~al.} (\bibinfo {collaboration} {HotQCD}),\ }\href
  {\doibase 10.1016/j.physletb.2019.05.013} {\bibfield  {journal} {\bibinfo
  {journal} {Phys. Lett. B}\ }\textbf {\bibinfo {volume} {795}},\ \bibinfo
  {pages} {15} (\bibinfo {year} {2019})},\ \Eprint
  {http://arxiv.org/abs/1812.08235} {arXiv:1812.08235 [hep-lat]} \BibitemShut
  {NoStop}%
\bibitem [{\citenamefont {Offler}\ \emph {et~al.}(2019)\citenamefont {Offler},
  \citenamefont {Aarts}, \citenamefont {Allton}, \citenamefont {Glesaaen},
  \citenamefont {J{\"a}ger}, \citenamefont {Kim}, \citenamefont {Lombardo},
  \citenamefont {Ryan},\ and\ \citenamefont {Skullerud}}]{Offler:2019eij}%
  \BibitemOpen
  \bibfield  {author} {\bibinfo {author} {\bibfnamefont {S.}~\bibnamefont
  {Offler}}, \bibinfo {author} {\bibfnamefont {G.}~\bibnamefont {Aarts}},
  \bibinfo {author} {\bibfnamefont {C.}~\bibnamefont {Allton}}, \bibinfo
  {author} {\bibfnamefont {J.}~\bibnamefont {Glesaaen}}, \bibinfo {author}
  {\bibfnamefont {B.}~\bibnamefont {J{\"a}ger}}, \bibinfo {author}
  {\bibfnamefont {S.}~\bibnamefont {Kim}}, \bibinfo {author} {\bibfnamefont
  {M.~P.}\ \bibnamefont {Lombardo}}, \bibinfo {author} {\bibfnamefont {S.~M.}\
  \bibnamefont {Ryan}}, \ and\ \bibinfo {author} {\bibfnamefont {J.-I.}\
  \bibnamefont {Skullerud}},\ }\href@noop {} {\bibfield  {journal} {\bibinfo
  {journal} {PoS}\ }\textbf {\bibinfo {volume} {LATTICE2019}},\ \bibinfo
  {pages} {076} (\bibinfo {year} {2019})},\ \Eprint
  {http://arxiv.org/abs/1912.12900} {arXiv:1912.12900 [hep-lat]} \BibitemShut
  {NoStop}%
\bibitem [{Note1()}]{Note1}%
  \BibitemOpen
  \bibinfo {note} {The factor $1/N_c$ in front of the gauge action was missing
  in Ref.~\cite {Aarts:2014nba}.}\BibitemShut {Stop}%
\bibitem [{\citenamefont {Morningstar}\ and\ \citenamefont
  {Peardon}(2004)}]{Morningstar:2003gk}%
  \BibitemOpen
  \bibfield  {author} {\bibinfo {author} {\bibfnamefont {C.}~\bibnamefont
  {Morningstar}}\ and\ \bibinfo {author} {\bibfnamefont {M.~J.}\ \bibnamefont
  {Peardon}},\ }\href {\doibase 10.1103/PhysRevD.69.054501} {\bibfield
  {journal} {\bibinfo  {journal} {Phys. Rev. D}\ }\textbf {\bibinfo {volume}
  {69}},\ \bibinfo {pages} {054501} (\bibinfo {year} {2004})},\ \Eprint
  {http://arxiv.org/abs/hep-lat/0311018} {arXiv:hep-lat/0311018} \BibitemShut
  {NoStop}%
\bibitem [{\citenamefont {L{\"u}scher}\ \emph {et~al.}(1996)\citenamefont
  {L{\"u}scher}, \citenamefont {Sint}, \citenamefont {Sommer},\ and\
  \citenamefont {Weisz}}]{Luscher:1996sc}%
  \BibitemOpen
  \bibfield  {author} {\bibinfo {author} {\bibfnamefont {M.}~\bibnamefont
  {L{\"u}scher}}, \bibinfo {author} {\bibfnamefont {S.}~\bibnamefont {Sint}},
  \bibinfo {author} {\bibfnamefont {R.}~\bibnamefont {Sommer}}, \ and\ \bibinfo
  {author} {\bibfnamefont {P.}~\bibnamefont {Weisz}},\ }\href {\doibase
  10.1016/0550-3213(96)00378-1} {\bibfield  {journal} {\bibinfo  {journal}
  {Nucl. Phys. B}\ }\textbf {\bibinfo {volume} {478}},\ \bibinfo {pages} {365}
  (\bibinfo {year} {1996})},\ \Eprint {http://arxiv.org/abs/hep-lat/9605038}
  {arXiv:hep-lat/9605038} \BibitemShut {NoStop}%
\bibitem [{\citenamefont {Chen}(2001)}]{Chen:2000ej}%
  \BibitemOpen
  \bibfield  {author} {\bibinfo {author} {\bibfnamefont {P.}~\bibnamefont
  {Chen}},\ }\href {\doibase 10.1103/PhysRevD.64.034509} {\bibfield  {journal}
  {\bibinfo  {journal} {Phys. Rev. D}\ }\textbf {\bibinfo {volume} {64}},\
  \bibinfo {pages} {034509} (\bibinfo {year} {2001})},\ \Eprint
  {http://arxiv.org/abs/hep-lat/0006019} {arXiv:hep-lat/0006019} \BibitemShut
  {NoStop}%
\bibitem [{\citenamefont {Edwards}\ and\ \citenamefont
  {Joo}(2005)}]{Edwards:2004sx}%
  \BibitemOpen
  \bibfield  {author} {\bibinfo {author} {\bibfnamefont {R.~G.}\ \bibnamefont
  {Edwards}}\ and\ \bibinfo {author} {\bibfnamefont {B.}~\bibnamefont {Joo}}
  (\bibinfo {collaboration} {SciDAC, LHPC, UKQCD}),\ }\href {\doibase
  10.1016/j.nuclphysbps.2004.11.254} {\bibfield  {journal} {\bibinfo  {journal}
  {Nucl. Phys. Proc. Suppl.}\ }\textbf {\bibinfo {volume} {140}},\ \bibinfo
  {pages} {832} (\bibinfo {year} {2005})},\ \Eprint
  {http://arxiv.org/abs/hep-lat/0409003} {arXiv:hep-lat/0409003 [hep-lat]}
  \BibitemShut {NoStop}%
\bibitem [{ope()}]{openqcd}%
  \BibitemOpen
  \href {{http://luscher.web.cern.ch/luscher/openQCD/}} {\emph {\bibinfo
  {title} {{OpenQCD, {\rm luscher.web.cern.ch/luscher/openQCD/}}}}}\BibitemShut
  {NoStop}%
\bibitem [{\citenamefont {Rantaharju}\ \emph {et~al.}(2018)\citenamefont
  {Rantaharju}, \citenamefont {Bennett}, \citenamefont {Dawson},\ and\
  \citenamefont {Mesiti}}]{Bennett:2018oyb}%
  \BibitemOpen
  \bibfield  {author} {\bibinfo {author} {\bibfnamefont {J.}~\bibnamefont
  {Rantaharju}}, \bibinfo {author} {\bibfnamefont {E.}~\bibnamefont {Bennett}},
  \bibinfo {author} {\bibfnamefont {M.}~\bibnamefont {Dawson}}, \ and\ \bibinfo
  {author} {\bibfnamefont {M.}~\bibnamefont {Mesiti}},\ }\href {\doibase
  10.22323/1.334.0039} {\bibfield  {journal} {\bibinfo  {journal} {PoS}\
  }\textbf {\bibinfo {volume} {LATTICE2018}},\ \bibinfo {pages} {039} (\bibinfo
  {year} {2018})},\ \Eprint {http://arxiv.org/abs/1806.06043} {arXiv:1806.06043
  [hep-lat]} \BibitemShut {NoStop}%
\bibitem [{fas()}]{fastsum}%
  \BibitemOpen
  \href {{http://fastsum.gitlab.io/}} {\emph {\bibinfo {title} {{FASTSUM
  collaboration, {\rm http://fastsum.gitlab.io/}}}}}\BibitemShut {NoStop}%
\bibitem [{\citenamefont {Glesaaen}\ and\ \citenamefont
  {J{\"a}ger}()}]{openqcd-fastsum}%
  \BibitemOpen
  \bibfield  {author} {\bibinfo {author} {\bibfnamefont {J.}~\bibnamefont
  {Glesaaen}}\ and\ \bibinfo {author} {\bibfnamefont {B.}~\bibnamefont
  {J{\"a}ger}},\ }\href {{https://doi.org/10.5281/zenodo.2216356}} {\emph
  {\bibinfo {title} {{openQCD-FASTSUM (v1.0), {\rm
  https://doi.org/10.5281/zenodo.2216356}}}}}\BibitemShut {NoStop}%
\bibitem [{\citenamefont {Glesaaen}()}]{openqcd-hadspec}%
  \BibitemOpen
  \bibfield  {author} {\bibinfo {author} {\bibfnamefont {J.}~\bibnamefont
  {Glesaaen}},\ }\href {{https://doi.org/10.5281/zenodo.2217028}} {\emph
  {\bibinfo {title} {{openQCD-hadspec (v0.1),\\{\rm
  https://doi.org/10.5281/zenodo.2217028}}}}}\BibitemShut {NoStop}%
\bibitem [{\citenamefont {Leinweber}\ \emph {et~al.}(2005)\citenamefont
  {Leinweber}, \citenamefont {Melnitchouk}, \citenamefont {Richards},
  \citenamefont {Williams},\ and\ \citenamefont {Zanotti}}]{Leinweber:2004it}%
  \BibitemOpen
  \bibfield  {author} {\bibinfo {author} {\bibfnamefont {D.~B.}\ \bibnamefont
  {Leinweber}}, \bibinfo {author} {\bibfnamefont {W.}~\bibnamefont
  {Melnitchouk}}, \bibinfo {author} {\bibfnamefont {D.~G.}\ \bibnamefont
  {Richards}}, \bibinfo {author} {\bibfnamefont {A.~G.}\ \bibnamefont
  {Williams}}, \ and\ \bibinfo {author} {\bibfnamefont {J.~M.}\ \bibnamefont
  {Zanotti}},\ }\href {\doibase 10.1007/11356462_4} {\bibfield  {journal}
  {\bibinfo  {journal} {Lect. Notes Phys.}\ }\textbf {\bibinfo {volume}
  {663}},\ \bibinfo {pages} {71} (\bibinfo {year} {2005})},\ \Eprint
  {http://arxiv.org/abs/nucl-th/0406032} {arXiv:nucl-th/0406032 [nucl-th]}
  \BibitemShut {NoStop}%
\end{thebibliography}%

\end{document}